\documentclass[12pt]{article}
\pdfoutput=1
\usepackage{epsfig,graphicx,amsbsy,amssymb,latexsym,amsfonts,amsmath}
\usepackage{pstricks,setspace}
\usepackage{hyperref,soul}
\usepackage{caption,subcaption}
\usepackage{eurosym}
\usepackage{color,tcolorbox}
\usepackage[a4paper]{geometry}
\usepackage{amscd}
\numberwithin{equation}{section}

\makeatletter
\newsavebox{\@brx}
\newcommand{\llangle}[1][]{\savebox{\@brx}{\(\m@th{#1\langle}\)}%
  \mathopen{\copy\@brx\kern-0.5\wd\@brx\usebox{\@brx}}}
\newcommand{\rrangle}[1][]{\savebox{\@brx}{\(\m@th{#1\rangle}\)}%
  \mathclose{\copy\@brx\kern-0.5\wd\@brx\usebox{\@brx}}}
\makeatother

\newcommand {\be}{\begin{equation}}
\newcommand {\ee}{\end{equation}}
\newcommand {\bea}{\begin{eqnarray}}
\newcommand {\eea}{\end{eqnarray}}

\newcommand{\C}{{\mathbb C}}

\newcommand{\I}{{\rm I}}
\newcommand{\IIstar}{{\rm II^*}}
\newcommand{\E}{{$\widetilde{\rm E}$}}
\newcommand{\F}{{\mathbb F}}

\newcommand{\ord}{{\rm ord\,}}
\renewcommand{\P}{{\mathbb P}}

\newcommand{\Tr}{{\rm Tr\,}}

\newcommand{\vol}{{\rm vol\,}}

\newcommand{\Z}{{\mathbb Z}}

\title{\bf  E(lementary)-Strings \\
in \\
Six-Dimensional Heterotic F-theory \\
}

\author{ \\
 \textsc{Kang-Sin Choi} $^{[1],[4]}$, \ \textsc{Soo-Jong Rey} $^{[2],[3],[4]}$
 \\
 }
\date{}
\begin{document}
\maketitle

\begin{center}
\renewcommand{\thefootnote}{\fnsymbol{footnote}} 
${}^{\thefootnote[1]}$ \sl Scranton Honors Program \\
Ewha Womans University, Seoul 03760 \rm KOREA \\[0.4cm]
${}^{\thefootnote[2]}$ \sl School of Physics and Astronomy \& Center for Theoretical Physics \\
          Seoul National University, Seoul 08826 \rm KOREA\\[0.4cm]
${}^{\thefootnote[3]}$ \sl Department of Basic Sciences \\
University of Science and Technology, Daejeon 34113 \rm KOREA \\ [0.4cm]
${}^{\thefootnote[4]}$ \sl Fields, Gravity \& Strings, {\rm CTPU} \\
Institute for Basic Sciences, Seoul 08826 \rm KOREA\\[1cm]
\end{center}

\begin{abstract}
Using E-strings, we can analyze not only six-dimensional superconformal field theories but also probe vacua of non-perturabative heterotic string. We study strings made of D3-branes wrapped on various two-cycles in the global F-theory setup. We claim that E-strings are elementary in the sense that various combinations of E-strings can form M-strings as well as heterotic strings and new kind of strings, called G-strings. Using them, we show that emissions and combinations of heterotic small instantons generate most of known six-dimensional superconformal theories, their affinizations and little string theories. Taking account of global structure of compact internal geometry, we also show that special combinations of E-strings play an important role in constructing six-dimensional theories of $D$- and $E$-types. We check global consistency conditions from anomaly cancellation conditions, both from five-branes and strings, and show that they are given in terms of elementary E-string combinations. \end{abstract}

${}$\\[500pt]

\tableofcontents

\section{Introduction and Summary}

Dynamics \cite{Witten:1995zh, Witten:1995gx, Witten:1996qb, Bershadsky:1996nh, Chou:1997ba} of NS5-branes \cite{Rey:1989xj, Rey:1989xi, Callan:1991ky, Rey:1991uu, Callan:1991dj} in type IIA string theory, equivalently, M5-branes \cite{Hull:1994ys, Witten:1995ex}
in M-theory has recently attracted renewed attention. Most important connections involve six-dimensional superconformal field theories (SCFTs) or six-dimensional little string theories with sixteen or eight supercharges \cite{Nahm:1977tg, Strathdee:1986jr}, and phase transition of small instantons of strongly coupled heterotic string theory \cite{Witten:1995zh,Witten:1995gx}. These studies enable us to understand intrinsically strong coupling dynamics of these theories on the 5-brane worldvolume in terms of geometry, and also open possibilities of wealthier non-perturbative vacua describing phenomena in our world \footnote{Other nontrivial situations involving strongly coupled six-dimensional superconformal field theories arise in the context of AdS/CFT correspondence \cite{Aharony:2015zea} and of M(atrix) theory \cite{Kim:1997gh}.} .

In studying them, various strings sourcing rank-2 self-dual tensor fields, forming tensor multiplets, play an important role in understanding worldvolume dynamics of such 5-branes. It was anticipated that multiple stack of the 5-branes have underlying non-Abelian symmetries of $ADE$-type. M-theory illustrates such configurations in elementary manner. M2-branes stretched between a pair of M5-branes or between M5-brane and M9-brane give rise to various strings, referred to as M-strings, first identified in \cite{Strominger:1995ac} and further studied in \cite{Distler:1996zg, Dijkgraaf:1996hk, Argyres:1996ak, Kim:1997uv}, and E-strings, first identified in \cite{Ganor:1996mu} and further studied in \cite{Seiberg:1996vs, Rey:1997hj, Minahan:1997ch, Minahan:1998vr}, which then combine with tensor multiplets to form non-Abelian multiplets. So far, except $A$-type theories, it has remained difficult to explicitly identify the underlying non-Abelian structure in terms of these building blocks. 

It is remarkable that, if we embed the 5-branes to dual F-theory \cite{Morrison:1996na, Morrison:1996pp, Morrison:2012np,Vafa:1996xn}, we may be able to see not only the emergence of $ADE$ symmetries but also far richer structure associated with them. With varying gauge coupling, the linearly aligned interval of M5-branes is now lifted to a non-trivially connected series of two-cycles in F-theory geometry. Essentially, this is the strategy that the six-dimensional ${\cal N} = (2,0)$ SCFTs are originally classified in relation to blown-up $ADE$ singularities in the type IIB string theory \cite{Witten:1995zh}. More recently, this structure of two-cycles can be accurately analyzed using algebraic geometric methods. This approach further made it possible to classify six-dimensional ${\cal N}= (1,0)$ SCFTs \cite{Heckman:2013pva,DelZotto:2014hpa}. 

In the F-theory side, the objects dual to M5-branes are not fundamental constituents; they are derived from intersections of singularities. We just recall that in F-theory every BPS objects are lifted to geometric singularities. It is due to the elliptic fibration which geometrizes varying axio-dilaton. So, discriminant loci of the elliptic fiber give rise to Kodaira surface singularities that are interpreted as 7-branes \cite{Vafa:1996xn}. If such two singularities collide, the resulting singularity becomes severer at the intersection and this leads to the effect that the gauge symmetry is locally enhanced. Usually, the resulting singularity at the intersection gives rise to extra massless matter from the enhanced gauge symmetry \cite{Katz:1996xe}. However, if the singularity is so severe that we have no gauge theory interpretation within the Kodaira classification, such intersection may be interpreted as a stack of 5-branes that is mapped to a stack of M5-branes in the M-theory side. We then blow up the singularity in the base, and the resulting exceptional two-cycle in the base would describe detachment of 5-brane \cite{Aspinwall:1997ye, Aspinwall:1996mn} (see also \cite{Morrison:2012np}). In particular, a D3-brane wrapped on the exceptional cycle gives rise to the M and E-strings attached to 5-branes. Understanding various building blocks for strings and 5-branes in M-theory is therefore translated to the analysis of possible two-cycles arising from blow-ups of enhanced singularity in F-theory. 

The F-theory compactified on K3 surface, which is an elliptic fibration over a $\P^1$, is dual to heterotic string theory compactified on a complex torus $\mathbb T$ \cite{Morrison:1996na}. With a single section of the elliptic fiber, we have two 7-branes harboring two respective $E_8$'s at the opposite `poles' of this base $\P^1$. We may further regard this $\P^1$ as a circle fibration over an interval $I$. We can then take fiberwise $T$-duality and obtain the heterotic M-theory on the interval $I$ \cite{Horava:1996ma}. These two 7-branes are mapped in M-theory to M9-branes (one of whose directions comes from the M-theory circle) at the ends of the interval $I$. 

Fibering this geometry on a common base, the duality can be extended to lower-dimensional spacetime. In particular, taking another $\P^1$ as a common base, one can construct nonperturbative six-dimensional heterotic string theories. In the heterotic description, a nontrivial vector bundle can be turned on and it is described by Yang-Mills instantons. In the F-theory description, these heterotic instantons are gauge symmetry enhancement points of $E_8$ \cite{Morrison:1996na, Morrison:1996pp} of the Hirzebruch base $ \F_n$. Again, these points are promoted to intersection points with another 7-brane $D'_{\rm inst}$. The beauty of F-theory is that we naturally obtain this $D'_{\rm inst}$ loci if we consider a compact internal manifold $X_3$ and decompose the discriminant. Namely, $X_3$ must satisfy the global consistency condition of vanishing first Chern class $c_1(X_3)$, and this condition is satisfied provided the discriminant loci are decomposed into two 7-branes describing $E_8 \times E_8$ gauge groups and $D'_{\rm inst}$ describing the heterotic instantons. 

Furthermore, by blowing up the base, we can transmute the instantons into 5-branes \cite{Morrison:1996na, Morrison:1996pp, Aspinwall:1997ye, Aspinwall:1996mn}.
In M-theory description, this is the so-called small instanton transition between Higgs branch of M5-branes dissolved inside the M9-branes and tensor branch of localized M5-branes in the heterotic interval $I$ \cite{Seiberg:1996vs}. Near the origin of this tensor branch, low-energy excitations involve a noncritical string, the M2-brane stretched between an M9-brane and M5-brane \cite{Ganor:1996mu}. This E-string describes fluctuation of the M5-brane relative to the reference location of M9-brane, as it becomes tensionless at the small instanton transition point.

In this paper, we ask what nonperturbative strings are made of. We investigate constituent strings of the heterotic F-theory and find that they include not only the above E-string and H-string (the original heterotic string) but also a variety of variant strings. By taking appropriate rigid and decoupling limits, we relate them to new structures of six-dimensional ${\cal N} = (1,0)$ SCFTs and little string theories. On the way, we also rederive constituent strings of the type II F-theory, viz. M-strings \cite{Haghighat:2013gba, Hohenegger:2013ala, Hohenegger:2015cba} and their affinization relevant for six-dimensional ${\cal N}=(2,0)$ SCFTs and  little string theories. Here, we highlight our main results. 

\begin{itemize}

\item An E-string is identified as a D3-brane (corresponding to an M2-brane in the M-theory side) wrapped on a two-cycle $E_p$ between the one 7-brane (corresponding to an M9-brane) and a 5-brane (corresponding to an M5-brane). 
We find that there are other types of E-string, depending on relative orientations with respect to the latter two objects. While the E-string stretch from the above 7-brane to a middle 5-brane, a D3-brane can also stretch from a middle 5-brane to the other 7-brane at the opposite pole, and we shall call it E$'$-string on equal footing to but distinguished from the E-string. There is also a conjugate to E-string, or \E-string for short, whose orientation along the cycle $E_p$ and the worldsheet are respectively opposite to that of E-string. They are BPS states preserving the same supersymmetries as the E-string. We also find that the worldsheet theory of the \E-string is formulated in terms of symplectic gauge group, rather than orthogonal gauge group for E-string. 

\item We find that that E, \E, and E$'$-strings are elementary constituents {(all of which we refer to as E(lementary)-strings)} in the sense that they generate all known string configurations. Besides the well-known bonding of E and E$'$-strings into heterotic string, $\rm E+\rm E' \rightarrow \rm H$ 
\cite{Haghighat:2014pva}, bonding E and \E-strings gives rise to M-string, $\rm E + \widetilde{E}  \rightarrow \rm M$. This is expected because M-strings parameterize the relative positions of 5-branes. We further find new strings as well, whose two-cycles are in the form $\rm E-E', E_1 + \frac{1}{2} (E_2 + E_3 + \dots ),$ required by algebraic structure or global structure. The 7-branes set absolute reference points and strings derived from them describe the relative motions of 5-branes.

\item The two-cycles associated with E-strings provide self-dual integral lattice. The standard root system generated by the lattice spans also $ADE$ lattices. In our heterotic F-theory setup, we can construct both simple and affine $ADE$ SCFTs. Conventional affine ${\cal N}=(2,0)$ SCFTs are obtained in the decoupling limit when, in the M-theory language, we zoom into a stack of M5-branes while separating M9-branes away. 

{
\item E(lementary)-strings enable to systematically analyze coalescence of small instantons and collision of singularities. The Higgs branch separates coalescent instantons into distinct points within the $E_8$ brane, while the tensor branch converts them into 5-branes and detach them into the bulk by blowing-ups. We can show that these operations on multiple instantons / 5-branes along the two branches do commute, and so the aforementioned SCFTs in terms of various strings discussed above can still be utilized in the analysis. This shall clarify the origins of 5-branes, only some of which are accounted for by the small instantons of the heterotic string theory.
}

\item Another important issue we are concerned through the paper is global consistency. It has been expected \cite{Vafa:2005ui} when global structure is taken into account, some bottom-up theories are not ultraviolet completed. If we take the internal geometry compact and keep the gravitational coupling finite, the low-energy vacua are subject to some global consistency condition. For this, we consider the above configurations that can be embedded in compact Calabi--Yau threefold, while relegating further aspects related to anomaly cancellations to a forthcoming paper \cite{choirey2}. 

\item It turns out that, not only the brane setup is constrained but also strings between 5-branes is necessarily modified for realizing theories of $D$ and $E$-type. The modification gives rise to a new kind of string with different global topology, taking the full canonical bundle into account. This makes other approaches difficult to analyze the dynamics of M5-branes. As M5-branes are induced from the colliding singularities, once the deformation of vacua are continuous, M5-brane configurations are always anomaly-free. We also find that in certain situations blow-up and extraction of 5-branes thereof are incompatible with maintaining the Calabi--Yau condition.

\item We shall see that two-dimensional anomaly on the string worldsheet can be cancelled if the sum of two-cycles becomes linearly equivalent to that of heterotic string. Ten-dimensional global consistency condition of heterotic string guarantees anomaly-free two-dimensional worldsheet theory. This enables us to classify anomaly-free configurations of M5-branes. The connection between E-strings and H-strings \cite{Haghighat:2014pva} from the viewpoint of anomalies is verified. We find it interesting to see that, as the self-duality of the tensor field it couples to already suggests so, M-string stretched between two M5-branes is chiral and contribute to anomalies. As an application, we may classify possible vacua of heterotic string theory including non-perturbative effect. 
 \end{itemize}

This paper is organized as follows. In Section 2, we recapitulate the six-dimensional setup of F-theory dual to heterotic string. We recall Kodaira's canonical singularities in the base and the corresponding 7-branes as well as small instantons. In Section 3, we take the base to be the Hirzebruch surface and show how the blow-up and blow-down of the surface enable us to identify various constituent strings, viz. E-strings, \E-strings, and E$'$-strings, all distinguished by their quantum numbers, and associated tensor multiplets. In Section 4, we will analyze how to build composites of the constituents. We will find that they can be systematically built by coalescing instantons and cycles. We also construct further constituent strings that are present in global models but decoupled in local limits. In Section 5, we use these results to construct global models whose matter contents are SCFTs or little string theories of $ADE$ types. We also present the prescription of decoupling the gravity and the noncritical strings.  In Section 7, we further study consistency conditions from both six-dimensional 5-brane worldvolume viewpoint and two-dimensional string worldsheet viewpoint. In Section 8, we discuss implications of our results to constructing new vacua. In the Appendix, we elaborate detailed analysis of Weierstrass equations for the emission and absorption of 5-branes and contrast local and global singularities.  

\section{F-Theory Setup}

In this section, we recapitulate compactification of F-theory dual to $E_8 \times E_8$ heterotic string \cite{Aspinwall:1997ye}. Upon further compactification on $\mathbb S^1$, this F-theory configuration is reduced to a heterotic M-theory configuration. Various M-branes in M-theory are mapped by duality to various geometric singularities in F-theory \cite{Aspinwall:1996mn}. 

We look for globally complete description in the sense that we keep the manifold compact and leave the configuration generic. We will see that all building blocks including M5-branes derived in this way satisfy global consistency conditions.  We will then be able to classify the resulting  theory of M5-branes in a systematic manner. On the way, we stress aspects that are unique to the global description. 
\subsection{F-theory dual to heterotic string}

Consider F-theory compactified on a Calabi--Yau threefold $X_3$. We take $X_3$ to be an elliptic fibration over a base surface $B$, $p:X_3 \to B$. We take the elliptic fiber to be defined by the Weierstrass equation
\be \label{Weier}
 y^2 = x^3 + f x + g,
\ee
where $f, g$ are complex coefficients that vary over the base surface $B$.  The curve is nonsingular provided the discriminant
\be \label{discriminant}
 \Delta = 4f^3 + 27 g^2
\ee
is nonzero \footnote{Twelve-dimensional supergravity description of this elliptic fibration is discussed in \cite{Choi:2014vya, Choi:2015gia}.} 

\begin{figure}[htbp]
 \begin{center}
\vskip-4cm
\makebox[\textwidth][c]{\includegraphics[angle=0,width=12cm]{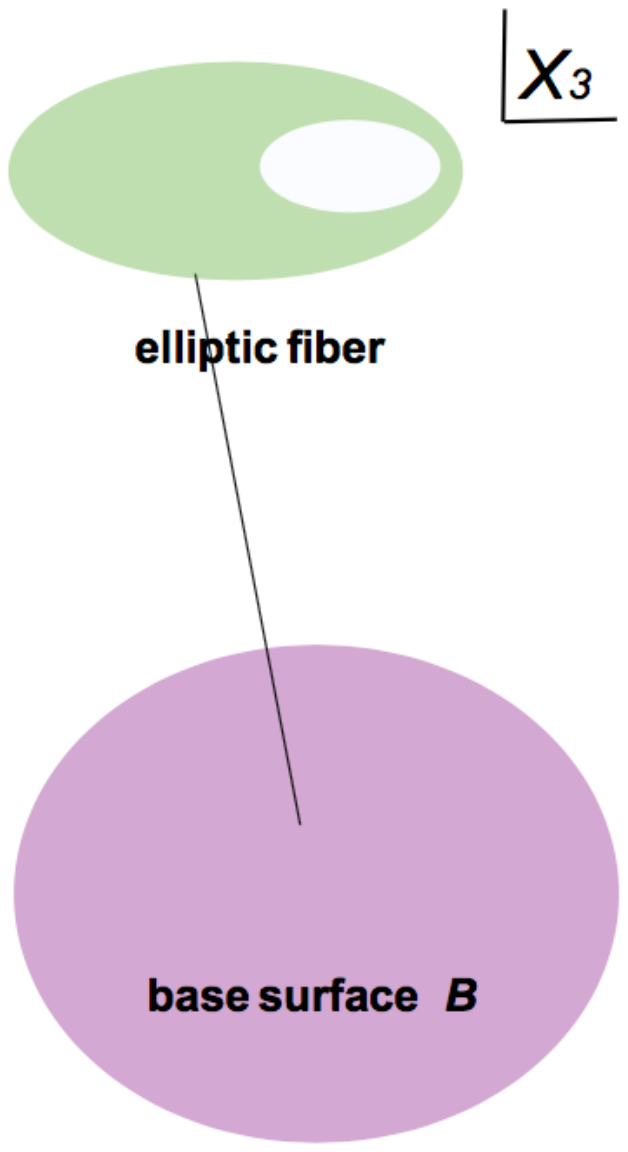}}
\end{center}
\vskip-4.5cm
\caption{\sl Elliptic fibration of the Calabi-Yau threefold $X_3$. The fiber may degenerate at the locus where the discriminant $\Delta$ of the Weierstrass equation vanishes.} \label{f:ellipticfibration}
\end{figure}

We assume that the fiber admits a global section. Homogeneity of the Weierstrass equation dictates that the line bundles $D,F,G$ associated with $\Delta,f,g$, respectively, should be powers of line bundles\footnote{We freely interchange the same notation for line bundles, their dual divisors, and the corresponding first Chern classes. The linear equivalence relation $\sim$ refers to co-homologous sections for the corresponding line bundles. \label{footnote:notations}} 
\be \label{powerRel}
 D \sim 12 {\cal L}, \quad F \sim 4 {\cal L}, \quad G \sim 6 {\cal L}.
\ee
We can fix the line bundle $\cal L$ from the fact that $X_3$ is a Calabi-Yau manifold.  
Namely, its first Chern class is known to have the dependence on the base as $ c_1(X_3) =- p^* ( K_B + {\cal L} ) $ and ought to vanish \cite{Morrison:1996na}. This asserts that the line bundle $\cal L$ is set by the canonical class $K_B$ of the base surface $B$,
\be \label{CYcond}
 {\cal L} = - K_B.
\ee

The elliptic fiber degenerates at the loci where the discriminant $\Delta$ vanishes. Their irreducible components within the base surface $B$ are classified according to the order of vanishing $\Delta, f, g$. Over each component, the Calabi-Yau threefold $X_3$ generically develops surface singularity of the type $\mathbb C^2 /\Gamma_G$, where $\Gamma_G$ is the quotient action of $ADE$ group $G$, whose algebra is in general non-simply-laced. Kodaira classified all possible canonical singularities of Weierstrass parametrization and the result is summarized in Table 1.

\begin{table}
\begin{center}
\begin{tabular}{ccccccc} \hline \hline
 ${\ord f}$ & ${\ord g}$ & $\ord\Delta$ & Kodaira & algebra &  $k$  \\ 
\hline
$\ge 0$ & $\ge 0$ & 0 & $\I_0$ & -&  -\\
$ 0$ & $ 0$ & $k \ge 1$ & $\I_k$ & $A_{k-1},C_{k}$ & -\\
$\ge 1$ & $1$ & $2 $& II & - & 1 \\
$1$ & $\ge 2$ & $3$ & III & $A_1$ &  -\\
$\ge 2$ & 2 & 4 & IV & $A_1,A_2$ &2 \\
$\ge 2$ & $\ge 3$ & 6 & $\I^*_0$ & $D_4,B_3,G_2$ & $3$ \\
$2$ & $3$ & $k \ge 7$ & $\I^*_{k-6}$ & $D_{k-2},B_{k-3}$ & - \\
$\ge 3$ &4& 8 & IV$^*$ & $E_6,F_4$ & $4$ \\
3 & $\ge 5$ & 9 & III$^*$ & $E_7$ & -\\
$\ge 4$ &5& 10 & $\IIstar$ & $E_8$ & $5$ \\
\hline
\end{tabular}
\end{center}
\caption{\sl Kodaira classification of singularities. The corresponding 7-branes supports the algebra, depending on further splitting or monodromy conditions \cite{Bershadsky:1996nh}. } \label{t:singularities}
\end{table}

We recall how these F-theory vacua can be constructed from string and M-theories via type II / F-theory duality. In type IIB string theory construction, the elliptic curve determines the type IIB dilaton-axion field configuration. It has generally nontrivial monodromy, and this monodromy is sourced by 7-branes wrapping around the singularity. It is now understood that, associated with each type of Kodaira singularity, the corresponding 7-brane configuration leads to enhanced gauge symmetry of $G$ type. In type IIA construction, the theory is compactified on $X_3$, the string coupling is taken infinite (thus becoming M-theory), and the area of elliptic curve is taken zero (thus decompactifying the F-theory circle). It is known that these geometric singularities in $X_3$ lead to enhanced gauge symmetry, determined by the monodromy of blow-down fiber components along cycles within the given component of the discriminant locus. 

\begin{figure}[h]
\begin{center}
\vskip-2cm
\begin{subfigure}{.5\textwidth}
\includegraphics[angle=0,width=14cm]{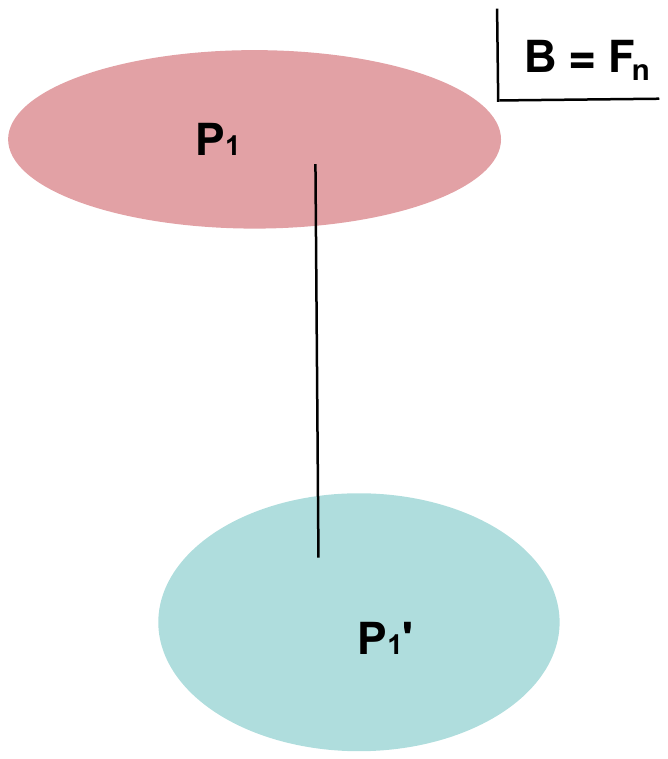}
\vskip-11cm
\caption{\sl The Herzebruch surface $\mathbb{F}_n$ of the base $B$.} \label{f:hirzbruch}
\end{subfigure}%
\begin{subfigure}{.5\textwidth}
\hskip-2cm
\includegraphics[angle=0,width=14cm]{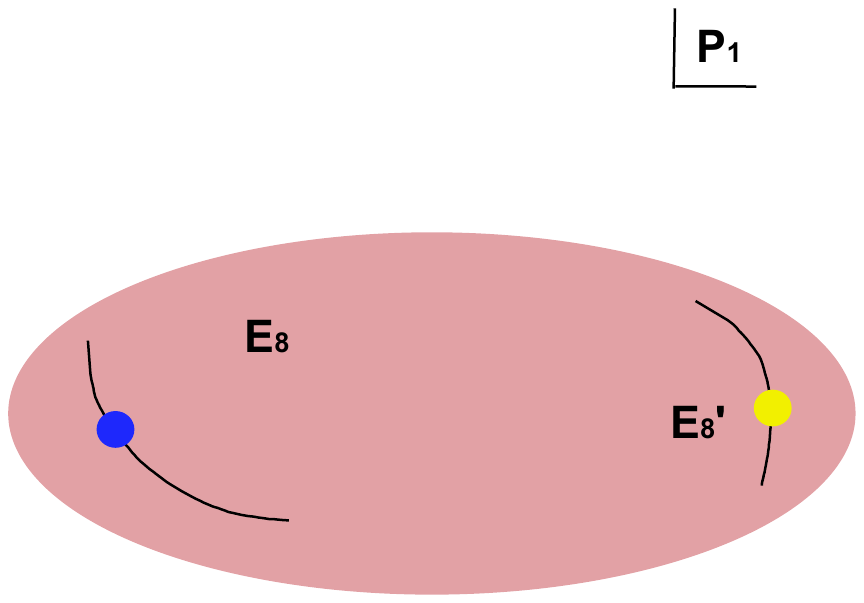}
\vskip-11cm
\caption{\sl Loci of seven branes for $E_8 \times E_8$ gauge structure.} \label{f:E8E8}
\end{subfigure}
\end{center}
\end{figure}

In this work, we are primarily interested in heterotic / F-theory duality. Our starting point is the eight-dimensional duality between F-theory on elliptic K3 surface and heterotic string theory on a complex torus $\mathbb T$. 
{The deformations of the Weierstrass equation span the moduli space\footnote{M(atrix) theory description of this heterotic T-duality was presented in \cite{Kabat:1997za}.} of heterotic string theory compactified on a complex torus $\mathbb T$,
\bea
{\cal M}_{\rm het}[\mathbb T] = {O(18, 2) \over O(18) \times O(2)}. 
\eea
}

{We extend the above duality fiberwise down to six dimensions.  On the heterotic side, we need Calabi--Yau twofold, so we should take the base to be $\P^1_{\rm b}$. This singles out the dual manifold $X_3$ to be K3 fibration over this $\P^1_{\rm b}$. As the K3 fiber is itself an elliptic fibration over a $\P^1_{\rm f}$, the base  surface $B$ of $X_3$ under elliptic fibration} is the Hirzebruch surface $\F_n$, a $\P^1_{\rm f}$ fibration over $\P^1_{\rm b}$ \footnote{Hereafter, we drop the subscripts, b(ase) and f(iber), as the reference should be clear within the contexts.} . It is described by the scaling equivalence in $\mathbb C^4$,
\begin{equation} \label{Hirzebruch} {
B = \left({\mathbb C}^4 - Z \right) / ({\mathbb C}^*)^2, \qquad ({\mathbb C}^*)^2}:
 (z',w',z,w)  \sim  (\lambda z', \lambda w', \mu z, \mu \lambda^n w).
\end{equation}
$(z', w')$ and $(z, w)$ span base and fiber of $\mathbb F_n$, respectively. 
In defining this equivalence, we are allowing negative values of degree $n$, and so we should exclude the curves $ Z=\{(z,w)=(0,0),(z',w')=(0,0)\}$. These curves form the Stanley--Reisner ideal $SR = \{ zw,z'w' \}$. 

The Hirzebruch surface $\mathbb{F}_n$ has two divisors spanning $H_{2}(\F_n)$: the zero section $\sigma = \{ z=0 \}$,   and the fiber  $f = \{z'=0\}$ satisfying the intersection relations \cite{GH} 
\be \label{Fnbasis}
\sigma \cdot \sigma= -n, \quad \sigma \cdot f = 1, \quad f \cdot f=0.
\ee
A redundant but useful divisor is the section `at infinity' $\sigma_\infty  = \{ w=0 \} \sim \sigma+n f$, which satisfies the intersection relations
\be
\sigma \cdot \sigma_\infty =0,\quad \sigma_\infty \cdot \sigma_\infty = n,\quad \sigma_\infty \cdot f = 1.
\ee
Its canonical class is given by 
\be \label{Fncanonical}
K_{\F_n} =- 2 \sigma -(n+2)f.
\ee  
On the heterotic side, we shall shortly see that we have small instantons whose number is related to the degree $n$. Moreover, a D3-brane wrapped on $f$ is identified with heterotic string \cite{Vafa:1996xn}.

\subsection{Seven-branes and small instantons}  \label{s:sevenbranes}

To fully explore the heterotic / F-theory duality,  we now introduce $E_8 \times E_8$ gauge structure. This is facilitated by the Kodaira type II$^*$ singularities having $ \ord (f,g,\Delta) = (4,5,10)$, as shown in Table \ref{t:singularities} \cite{Morrison:1996na,Morrison:1996pp,Aspinwall:1996mn}. At the singularities, there are ten 7-branes, of which eight are D7-branes connected by fundamental strings and two are certain $(p,q)$ 7-branes connected by string junctions, spanning the $E_8$ root lattice. 
When these singularities are located at $\sigma$ and $\sigma_\infty$, the Weierstrass equation reads
\begin{equation} \label{smallinst}
 y^2 = x^3 +  \left( f_8 (z',w') z^4 w^4 \right) x + \left( g_{12-n}(z',w') z^5 w^7+  g_{12}(z',w') z^6 w^6 + g_{12+n}(z',w') z^7 w^5 \right),
\end{equation}
where $z, w$ {and $z', w'$} are affine coordinates of two $\mathbb P^1$'s in $X_3$, respectively.

The corresponding discriminant loci, $z=0$ and $w=0$, look like two points in the fiber $f$. So long as generic $f_8$ and $g_{12}$ in Eq.(\ref{smallinst}) are allowed, these two loci are interpreted to support 7-branes having $E_8$ gauge group on each eight-dimensional world-volume \cite{Morrison:1996na,Morrison:1996pp}. The zeros of $g_{12-n}$ and $g_{12+n}$ are locations of small instantons \cite{Witten:1995gx,Morrison:1996pp}. At these locations, the singularities are worsened, reflecting the fact that gauge transformations are singular.

Deformations of the Weierstrass equation (\ref{smallinst}) result in milder singularity and the gauge symmetry supported at the singularity becomes smaller. Because of this property, such deformations are regarded as Higgsing. {In what follows, we limit our study to geometric deformation.\footnote{For non-geometric deformation, see \cite{McOrist:2010jw,Font:2016odl,Malmendier:2014uka,Gu:2014ova}.}} The most generic form of deformation is provided by \cite{Morrison:1996na,Morrison:1996pp}
\begin{equation} \label{generic}
 y^2 = x^3 +\left(  \sum_{k=-4}^4 f_{8+kn}(z',w') z^{4+k} w^{4-k} \right) x + \left( \sum_{l=-6}^6 g_{12+ln}(z',w') z^{6+l} w^{6-l} \right).
\end{equation}
From the intersection relations Eq.(\ref{Fnbasis}), we deduce that $z,z'$ are respective sections of the bundles ${\cal O}(-n)$ and ${\cal O}(1)$ over the base $\sigma$. Thus, the coefficients $f_{8+kn}$ and $g_{12+ln}$ are polynomials of degree 
\be
 (F - (4+k) \sigma) \big|_\sigma = 8+kn \qquad \mbox{and} \qquad (G - (6+l) \sigma) \big|_\sigma = 12+ln, 
\ee
respectively. The allowed deformations are spanned by monomials with non-negative degrees. So, requiring that $f, g$ having negative subscript coefficients are vanishing, even for generic deformation as in Eq.(\ref{generic}), we have different orders of $f,g$ and $\Delta$ for different $n$. This is the maximal Higgsing we can perform \cite{Morrison:2012np}. Note that the self-intersection of the $\sigma$ is $-n$, as given in Eq.(\ref{Fnbasis}). Conversely, if we take an effective irreducible $\P^1$ divisor in the base with the self-intersection $-n$, we can determine the resulting gauge group for the maximal Higgsing \cite{Morrison:2012np}.

The discriminant $\Delta$ in Eq.(\ref{discriminant}) is factorizable into components. That is, the corresponding divisor is decomposed as
\begin{equation} \label{disclocus}
 D =  \sum (\ord \Delta)_a D_a + D'.
\end{equation}
 We assume that every divisor $D_a$ is effective, irreducible, and supports canonical singularity of Kodaira classification, including the smooth II. Each divisor $D_a$ is then interpretable as a locus on which a stack of 7-branes are wrapped. 
The residual part is denoted as $D'$. The number of 7-branes in each stack is the order of the discriminant $\Delta$ evaluated at $D_a$, which is the order of discriminant at the singularity. Likewise, we can decompose $F$ and $G$ according to Kodaira's classification with $\ord f,\ord g$ evaluated at $D_a$, respectively, \cite{Aspinwall:1996mn} and call the residual part as $F'$ and $G'$.
Physically, these relations are understood as the extension of local charge conservation to compact spaces, resulting in the so-called `global consistency conditions' \cite{Sadov:1996zm}.
 
Here, we are considering $E_8\times E_8$ gauge group and its deformations. That is, all the terms in Eq.(\ref{smallinst}) should be present in Eq.(\ref{generic}), for which we require
\be
 \ord f|_\sigma \ge 4, \quad  \ord g|_\sigma \ge 5,
\ee
and similar for those evaluated at $\sigma_\infty$.
These conditions are lifted to those of divisors
\be
 F' \ge 4f, \quad G' \ge 5f
\ee
to be generic.
Also, the locations of small instantons are promoted to the intersections of $\sigma$ and $\sigma_\infty$ with the following divisors \cite{Aspinwall:1997ye}
\begin{equation} \label{instantonic7} \begin{split}
 D' _\text{inst}& \equiv D - 10 \sigma - 10 \sigma_\infty = 4 \sigma + (2n+ 24) f, \\
 F'_\text{inst} & \equiv F - 4 \sigma - 4 \sigma_\infty = 8f, \\
 G'_\text{inst} & \equiv G - 5\sigma - 5 \sigma_\infty = 2 \sigma + (n+12) f.
\end{split}
\end{equation}
Indeed, they give the locations of small instantons
\begin{align}
D' _\text{inst}|_{\sigma} &=  2(12-n), \quad G'_\text{inst} |_\sigma =12-n \label{E8inst} \\
 D' _\text{inst}|_{\sigma_\infty} & = 2(12+ n), \quad G'_\text{inst}|_{\sigma_\infty} = 12+n \label{E8primeinst},
\end{align}
where the factor 2 is due to squaring $\Delta \simeq g_{12 \pm n}^2$. It means that the divisor $D' _\text{inst}$ hits $\sigma$ and $\sigma_\infty$, respectively, by $(12-n)$ and $(12+n)$ times cuspidally \cite{DelZotto:2014hpa},\cite{Aspinwall:1997ye}. 

Deformations as in Eq.(\ref{generic}) make the singularity at $z=0$ milder, corresponding to Higgsing $E_8$ to a lower rank gauge group. Still, the small instantons are controlled by the coefficients $g_{12-n}$ of $z^5 w^7$ and $g_{12+n}$ of $z^7 w^5$. In other words, the small instanton singularities are governed by the residual part $D'_{\rm inst}$. Thus, the deformations (\ref{generic}) is naturally interpreted as instantons growing into finite sizes in the heterotic side. 

The residual part $D'$ in Eq.(\ref{disclocus}), now different from $D'_{\rm inst}$, contains the information of the hypermultiplet matter contents. For instance, we may have deformation terms for $E_8$ in Eq.(\ref{smallinst})
\be
 f_{8-n} z^3 w^5 x + g_{6-n}^2 z^4 w^{8}.
\ee
Here, we still interpret that we have $E_6$ at $z=0$. In the heterotic side, we still have $(12-n)$ small instantons at $z=0$ since the coefficient $g_{12-n}$ is nonzero. Some of the instantons have finite size and are embedded in $SU(3)$ structure group that is the commutant of $E_6$ in $E_8$. We are decomposing the discriminant as 
$$ D = 8 \sigma + 10 \sigma_\infty + (6 \sigma + (2n+24)f), $$
where we identify the last term in the parentheses as $D'$. Then $D'|_\sigma = 4(6-n)$ gives the information about the $(6-n)$ matter curves for $\bf 27$ of $E_6$ in the hypermultiplets, with splitting multiplicity 4 \cite{Morrison:1996pp,Bershadsky:1996nh,Choi:2009tp}.

\section{Branes and E-Strings}
We next move to identify branes and E-strings in the F-theory setup dual to heterotic string theory. We begin with analysis on small instanton points for the simplest case when all the zeros of $g_{12-n}(z')$ in Eq.(\ref{smallinst}) are distinct. Without loss of generality, we may take one of them at $z' =0$, or 
\begin{equation} \label{smallinstemitted}
g_{12-n}(z',w') = g_{11-n}(z',w') z',
\end{equation}
while $g_{11-n} (0,w') \ne 0$.
{\em Locally,} at  $z'=0$, we have Kodaira II singularity $y^2 \simeq x^3 + z' z^5$, for ord$(f,g,\Delta)=(\infty,1,2)$ \footnote{Recall that the order of singularity is zero  if we have the constant or infinite in the absence of the $z'$-dependent term.} \cite{DelZotto:2014hpa}.  It means that the discriminant  behaves approximately as $(z')^2$ around $z'=0$. However, the full discriminant does not vanish at this point. We will come to this point later.
This situation is also described by the observation that $D'_\text{inst}$ in Eq.(\ref{instantonic7}) intersects $\sigma$ as in Eq.(\ref{Fnbasis}), which localizes the $\IIstar$ singularity at $z=0$, at 
\be \label{p}
 p \equiv \{z= z'=0\} \subset \sigma \cap D'_\text{inst}.
\ee 

From now on, we will blow up and blow down the singularity and identify all possible E-strings. 

\subsection{Blow up}

At the intersection point $p$ in Eq.(\ref{p}), the two singularities $\I\I$ and $\IIstar$ collide. The resulting singularity becomes severer than those classified by Kodaira, for $\ord(f,g,\Delta)$ are greater than $(4,6,12)$. If this were a surface singularity, then we cannot have appropriate sections for the elliptic fiber making the first Chern class of $X$ vanish as in Eq.(\ref{CYcond}) \cite{Aspinwall:1996mn,Morrison:2012np}.  However, in our case it is a curve singularity that can be smoothened out by resolving only in the base and the accompanying proper transforms automatically satisfy the condition. The resulting process describes the emission of a 5-brane from $p$, which is dual to M5-brane.

Consider blowing up at $p$ in the base $ \pi: \F_n^{(1)} \to \F_n$ \cite{Aspinwall:1997ye}. The proper transforms are
\begin{align} 
\pi^* \sigma & \sim \sigma' + E_p,  \label{blow-upbase} \\ 
\pi^* f_p & \sim E'_p + E_p, \label{semistabledeg}
\end{align}
where $E_p$ refers to the exceptional divisor isomorphic to $\P^1$. 
The associated two-form becomes a new element of $H^{1,1}(\F_n^{(1)})$. Also, the canonical class Eq.(\ref{Fncanonical}) gets modified, as
\be \label{ModCanonical}
 K_{\F_n^{(1)}} = \pi^*K_{\F_n} + E_p = -2 \pi^* \sigma -(2+n) \pi^*f + E_p.
\ee

\begin{figure}[h]
\begin{center}
\vskip-2cm
\includegraphics[angle=0,width=14cm]{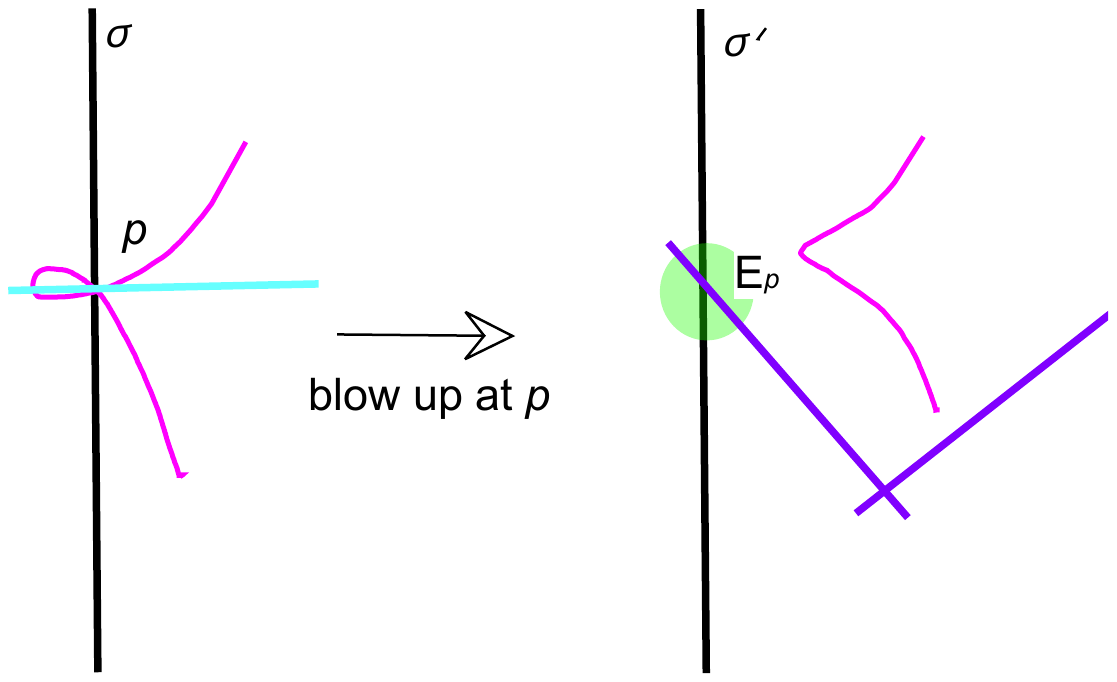}
\end{center}
\vskip-13cm
\caption{\sl The location of small instanton $p$ is the intersection between the discriminant loci $D_{\rm inst}'$ and $\sigma$. Blow up at this point $p$ in the Hirzebruch base $\F^{(1)}_n \to \F_n$ ejects the small instanton into the bulk. The curve $\sigma$ is now transformed into $\sigma'$ and $E_p$.} \label{f:blow-up}
\end{figure}

The exceptional divisor $E_p$ has self-intersection $(-1)$ and intersects $\sigma'$ at one point
\be \label{EpdotSigma}
 E_p \cdot E_p=-1, \quad E_p \cdot \sigma' =1.
\ee
Careful analysis of the Weierstrass equation, as given in the Appendix A, shows that $E_p$ is not a discriminant locus but locally support the type II Kodaira fiber. Therefore, there is no gauge theory supported at the 7-brane on $E_p$.

After the resolution, $\sigma'$ replaces the role of the original $\sigma$ supporting the $E_8$. Using the relations (\ref{blow-upbase}) and (\ref{EpdotSigma}), one can verify that 
\be 
\label{SigmaprimeIntersec}
\sigma' \cdot \sigma' = -(n +1).
\ee 
This implies that the $E_8$ on $\sigma'$ carries only $12-(n+1)=11-n$ small instantons, and so one instanton must be ejected out.
At places where Eq.(\ref{smallinstemitted}) is finite, the fiber is non-degenerate and remains irreducible as before. We denote it by the same name $f$, abbreviating pullback by $\pi$.  Indeed, the fiber $f$ still intersects the $\sigma'$ once, viz. $\sigma' \cdot f = 1$. 

We also have $E'_p$ as the proper transform of $f_p$ in Eq.(\ref{semistabledeg}). It is not an exceptional divisor, but it still has properties similar to those of $E_p$. Indeed, $E'_p$ has the same self-intersection number,
\be \label{Castelnuovo}
 E_p' \cdot E_p' = (f_p - E_p)\cdot (f_p - E_p) = -1 
\ee 
because $f_p \cdot E_p = 0$, and also intersects $\sigma_\infty$ once:
\be \label{equalrel}
 \quad E_p' \cdot \sigma_\infty = (f_p -E_p) \cdot  (\sigma + n f_p)= 1.
\ee
So, one can say that the fiber $f_p$ over $p$ underwent degeneration into two equal divisors, $E_p$ and $E'_p$  \cite{Aspinwall:1997ye}. They meet at one point transversally 
\be \label{M5location}
 E_p \cap E_p' \equiv p_t.
\ee
If we go to M-theory by shrinking one of the cycles of elliptic fiber, the worldvolume at this intersection point $p_t$ is mapped to an M5-brane. As such, we shall refer to this intersection as a `5-brane'.

\begin{figure}[h]
\begin{center}
\includegraphics{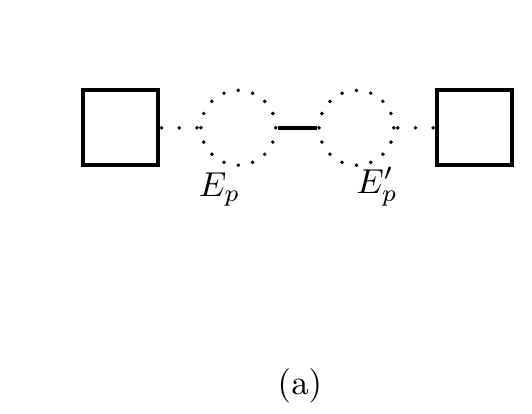}
\qquad
\includegraphics{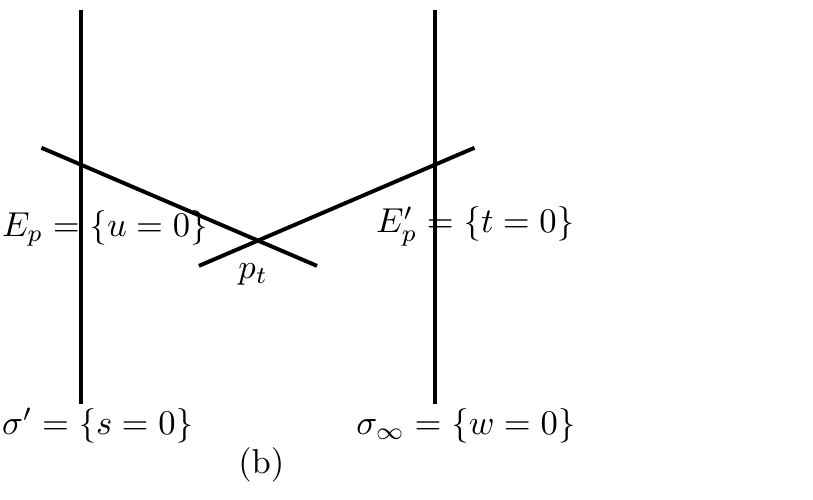}
\end{center}
\caption{ \sl  (a) The Dynkin diagram for elementary E-strings. Dotted nodes denote the $\P^1$ cycles, on which a wrapped D3-brane yields an E-string. The lines denote the intersection between cycles. In particular, the solid line is to be interpreted as an M5-brane. The boxes are cycles supporting $E_8$ symmetries.  (b) A dual graph highlights the intersection structure of these cycles, which are denoted by lines. The coordinates will be used in the Appendix.} \label{f:basicE}
\end{figure}

In general, it can happen that the modified canonical class (\ref{ModCanonical}) of the new base does not satisfy the Calabi--Yau condition (\ref{CYcond}). After proper transforms, we can rescale the coordinates and parameters $x \to u^2 x,y \to u^3 y, f\to u^4 f,g \to u^{6} g$ so as to modify the line bundle $\cal L$ by an $E_p$ associated with the $u$:
\be \label{modificationL}
 {\cal L} \to {\cal L} - E_p. 
\ee
Due to the relation (\ref{powerRel}) of line bundles involved, such scaling is possible only at a surface singularity that is severer than $\ord(f,g,\Delta) = (4,6,12)$. Geometrically, any point in the base can be blown up, yet this does not lead the resulting space $X_3$ to be a Calabi--Yau threefold satisfying the condition (\ref{CYcond}). For instance, blow-up at a point of $\sigma$ where there is no small instanton still changes the base as in Eq.(\ref{SigmaprimeIntersec}), but it ruins the Calabi-Yau condition. 

Also, in the global case, this scaling ought to hold for the entire equation (\ref{Weier}), not just keeping relevant terms. For instance, the intersection between $D_{\rm inst}'$ and $\sigma$ is possible. However, consistency at other places than $z=0$ is not guaranteed. That is, we may blow-up in the bulk of $B$, but the Calabi--Yau condition (\ref{CYcond}) is ruined.

\subsection{Blow down}
We now move to blow down. Note again that the divisor $E_p'$, which is the proper transform of $f_p$ as in Eq.(\ref{semistabledeg}), has also self-intersection $(-1)$, as in Eq.(\ref{Castelnuovo}). By the Castelnuovo criterion \cite{GH}, we may blow it down to obtain the $\F_{n+1}$ base, and the situation becomes heterotic string with $(11-n)$ and $(13+n)$ small instantons. Indeed, it agrees with the canonical class
\be
 K_{\F_n^{(1)}} = -2 \sigma - (2+n) f + E_p = -2 \sigma - (2+(n+1)) f + E_p'  = K_{\F_{n+1}} + E_p'.
\ee
We have also seen in Eq.(\ref{SigmaprimeIntersec}) that the self-intersection of $\sigma'$ is $-(n+1)$. Thus, the resulting base is indistinguishable from $\F_{n+1}$, which is a blow-up at the location of small instanton $ z'=w=0$ on the `right' $E_8$ at $\sigma_\infty$, modulo possible coordinate redefinitions. So, in this case, if we rename the exceptional divisor as $E_p'$ and the proper transform of $f_p$ as $E_p$, everything becomes identical as before. 
Beside notational asymmetry, we have democracy for exchanging $E_p$ and $E_p'$. 

Note again the linear equivalence relation (\ref{semistabledeg}). 
Although $E_p$ and $E_p'$ intersects at a point, the  summed divisor $f_p$ is again a $\P^1$ fiber over the $p$, parameterizing the separation of two $E_8$ branes at $\sigma$ and $\sigma'$. 
Therefore, in this global F-theory with heterotic duality, the {\em absolute} location of the 5-brane can be parameterized by either $E_p$ or $E'_p$ from the respective reference $\sigma'$ and $\sigma_\infty$. 

Putting together the analysis, we conclude that canonical singularities can be blown up and down in a sequential and continuous manner. 

\subsection{E-strings} \label{sec:Estring}

In this section, we introduce and analyze E-string in F-theory. We eventually find that, in the contexts of global geometry, we have to deal with not only E-string but also other variant strings of it.

\subsubsection*{E-string}

A D3-brane wrapped on the two-cycle $E_p$ in Eq.(\ref{blow-upbase}) yields an E-string. Its tension is proportional to the volume of $E_p$. We are particularly interested in its collapsing limit. As $E_p$ intersects $\sigma'$ that supports the $\IIstar$ ($E_8$) singularity, the E-string is charged under the $E_8$ symmetry. In the global description, $E_8$ is gauge symmetry, whose interaction coupling strength is inversely proportional to the volume of $\sigma'$. In local description, we send the volume to infinity and the $E_8$ becomes a flavor symmetry, the setup studied in \cite{Ganor:1996mu, Seiberg:1996vs}. The other end of the E-string touches the 5-brane at $p_t$ in Eq.(\ref{M5location}), which is mapped to M5-brane in the M-theory.

On the worldsheet of E-string, we have two-dimensional ${\cal N}=(0,4)$ supersymmetry. We can identify symmetries of the system most manifestly in the M-theory side, where the E-string is the boundary of M2-brane. The four supersymmetry generators transform as $\bf (2,1,2,1)_-$ under the symmetry
\be \label{Rsymmetry6D}
SU(2)_L \times SU(2)_R \times SU(2)_G \times SU(2)_D \subset  Spin(4)_\parallel \times Spin(5)_{\sf R},  
\ee
which is the rotational symmetry of boundary of M2-brane in the $\mathbb C^2 \times \mathbb C^2$ subspace of both inside and transverse to the M5-branes. In particular, the $R$-symmetry follows geometrically from $SU(2)_L \times SU(2)_G \simeq SO(4)$ \cite{Gadde:2015tra}. The subscript ``$-$'' signifies the $SO(1,1)$ chirality on the worldsheet. The winding number $q$ of the D3-brane becomes the charge of the two-form $\omega_p$ dual to the $E_p$ \footnote{So, the condensation of these D3-branes triggers the Higgs mechanism for the corresponding two-form fields \cite{Rey:1989ti}. }. The localized zero modes on the E-string are \cite{Kim:2014dza}
\begin{itemize}
 \item $O(q)$ symmetric hypermultiplet $\bf (q+2)(q-1)/2+1$: scalars parameterizing the collective motion of D3 within the 5-brane and fermions $({\bf 1,2,2,1})_-$, 
 \item $O(q)$ antisymmetric vector multiplet $\bf q(q-1)/2$: gauge boson and gaugino $({\bf 2,1,2,1})_+$, and
 \item $E_8$ adjoint Fermi multiplet: sixteen Majorana--Weyl fermions for the $E_8$ current algebra, neutral under all the tangent and normal bundles, localized at the intersection between D3 and 7-brane on $\sigma'$.
\end{itemize}

This worldsheet gauge theory is best analyzed in the dual theories \cite{Kim:2014dza}. More specifically, in M-theory, the $\IIstar$ singularity at $\sigma$ is mapped to M9-branes and D3-brane to M2-brane in the M-theory. Further compactifying on a small circle, these become O8/D8-branes in type I$'$ string theory, imposing boundary condition on the worldvolume fields on D2-brane that ends on them \cite{Gaiotto:2008sa}.

\subsubsection*{$\bf \widetilde E$-string}
We also have a string conjugate to E-string, obtained by a D3-brane wrapped on $-E_p$. For short, we refer to it as \E-string.  Although the cycle $-E_p$ has the opposite orientation to $E_p$, if the resulting string along the remaining directions has the opposite orientation as well, then the D3-brane is the same BPS state as that of E-string, preserving the same worldvolume supersymmetry. In fact, in the limit where the 7-brane becomes an O7-brane, the combination of the reflection of the transverse directions and the worldsheet orientation reversal is a symmetry of the string theory. The resulting string is negatively charged source to the dual two-form to $\omega_p$. This D3-brane sees 5-brane in the same way as the E-string does, while leaving the directions parallel to 5-brane intact. Thus, all the fields on the \E-string are related to those on the original E-string by flipping the $SU(2)_G \times SU(2)_D$ chirality \cite{choirey2}
\be \label{minusEexchange}
 E_p \leftrightarrow -E_p: \quad q \leftrightarrow -q, \quad SU(2)_G \leftrightarrow SU(2)_D, \quad \text{symmetrization} \leftrightarrow \text{antisymmetrization}
\ee
Due to the opposite orientation of D3, all the boundary conditions on the worldsheet field acquire the extra minus sign, so that we have opposite projection by the orientifold {\cite{Gaiotto:2008sa}}. The worldsheet gauge group is now $USp(q)$. The resulting field theory is a new quiver gauge theory with the following field contents;
\begin{itemize}
 \item $ USp(q)$ antisymmetric hypermultiplet $\bf (q-2)(q+1)/2+1$: scalars for the motion of D3 within the 5-brane and fermions $({\bf 1,2,1,2})_-$. 
 \item $ USp(q)$ symmetric vector multiplet $\bf q(q+1)/2$: gauge boson and gaugino $({\bf 2,1,1,2})_+$, and
 \item $E_8$ adjoint Fermi multiplet: localized sixteen chiral fermions at the D3 and 7-brane intersection.
\end{itemize}

Our convention is $USp(2)=SU(2)$. Recall that, in the classification of semisimple Lie group, the $USp(q)$ with odd integer $q$ is absent. However we may define it as a group leaving the bilinear form $J=H \oplus H \oplus  \dots \oplus H \oplus 0$ invariant, where $H=(\begin{smallmatrix} 0 &1 \\ -1 & 0\end{smallmatrix})$ is the hyperbolic form, relaxing the non-degenerate condition \cite{Proctor}.\footnote{The discussion of $USp(1)$ from orientifold planes can be found in \cite{Opminus}.}

We have an interesting consistency check. It is known that we may define $SU(-q)$ group of negative rank ($q \in {\mathbb N}$) by exchanging symmetrization and antisymmetrization in tensoring representations of $SU(q)$ \cite{Cv}. As different projections, we also formally define $O(-q) \equiv USp(q)$ with $q \in {\mathbb N}$ and vice versa by the same kind of exchange \cite{King,Rey:1997gc}. 

\subsubsection*{E$'$-string and $\bf \widetilde E'$-string}
We also have another two-cycle $E'_p$ connected to $\sigma_\infty$ supporting the other $E_8$, as in Eq.(\ref{equalrel}). A D3-brane wrapped on this cycle gives another type of, let us call, E$'$-string. 
This string sees relatively opposite orientation to 5-brane. Thus, we have  $SU(2)_L \leftrightarrow SU(2)_R$. On the other hand, the orientation of the D3-brane remains the same, in effect yielding 
\be \label{Eprimeexchange}
 E_p \leftrightarrow E_p': \qquad SU(2)_L \leftrightarrow SU(2)_R, \qquad SU(2)_G \leftrightarrow SU(2)_D, \quad
 E_{8, L} \leftrightarrow E_{8, R}.
\ee
Therefore, when two strings of E and E$'$ meet at a common intersection along 5-brane, the two-dimensional worldsheet supersymmetry is locally enhanced to $(0,8)$, supplemented by generators $\bf (2,1,1,2)_-$ \cite{Hohenegger:2015cba}. This is the same supersymmetry preserved by heterotic string for which two strings are merged \cite{Polchinski:1995df}.

Finally, we have \E$'$-string that is obtained by a D3-brane wrapped on the cycle $-E_p'$ and flipping orientation of the remaining worldsheet. An E-string is converted to \E$'$-string also by the successive operations (\ref{minusEexchange}) and (\ref{Eprimeexchange}).

\section{Singularity Enhancement and M-strings} \label{sec:Mstrings}

The worldvolume theory from a stack of coincident M5-branes is known to be six-dimensional ${\cal N} = (2,0)$ {SCFT} having nonabelian structure, admitting $ADE$ classification \cite{Witten:1995zh}. Their fluctuation is translated to dynamics of M-strings \cite{Haghighat:2013gba}. If such 5-branes arise from blowing up small instanton points that can be probed by E-strings, M-strings can also be understood as combinations of E-strings. { In Sections \ref{sec:Mstrings}.1, we consider such possibility in the simplest case where we blow up two small instanton points. This leads us to rediscover $A_1$ SCFT and M-strings in terms of E-strings in Section \ref{sec:Mstrings}.2. Moreover, considerations of global embedding in Section \ref{sec:Mstrings}.3 and conjugate two-cycles in Section \ref{sec:Mstrings}.4 bring us to identify new kind of strings.}

\subsection{Two instantons}

\begin{figure}[t] 
\begin{center}
\includegraphics{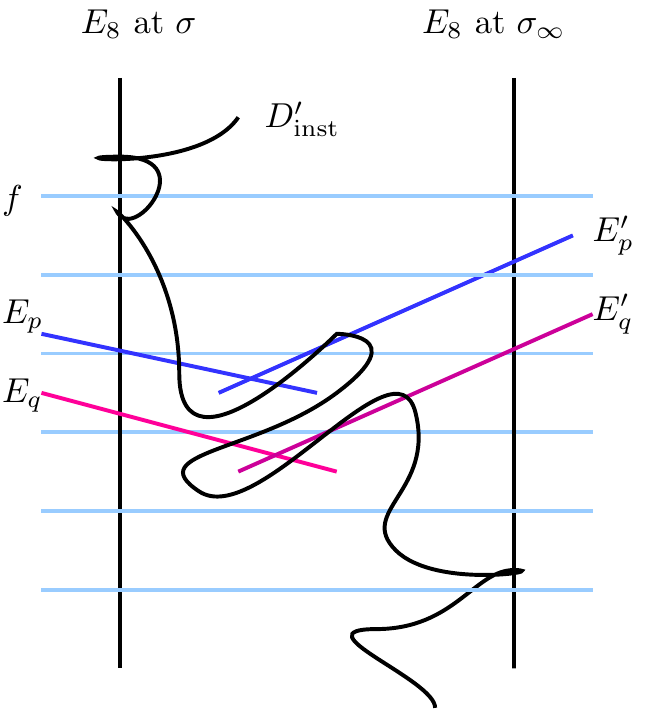}
\caption{\sl Two emitted instantons. All the lines in the figure denote $\P^1$, highlighting intersection structure. Vertical lines are bases locally supporting II$^*$ curves, while horizontal lines are fibers of Hirzebruch surface. Blow-up at some two locations of small instantons $\{p,q\} \subset \sigma$ give rise to exceptional divisors $E_p$ and $E_q$.  \label{f:twoEs}} 
\end{center}
\end{figure}

On top of the blow-up $\pi:\F_n^{(1)} \to \F_n$ at $p$ in Eq.(\ref{smallinstemitted}), 
we may proceed to detach another small instanton located at another point $q \in \sigma$. Under $\pi$, that point $q$ ($ \ne p$) is mapped to a point in $\sigma'$, which again shall be referred to as $q$ without confusion. We now perform another blow-up at $q$ to have $\F_n^{(2)} \to \F_n^{(1)}.$ The fiber  $f_q$ passing through $q$ degenerates in the same way,
\begin{align} \sigma' &\sim \sigma'' + E_q, \\ 
f_q &\sim E_q'+E_q, \\
K_{\F_n^{(2)}}&=- 2 \sigma - (n+2) f +E_p + E_q, \label{canonicalttF}
\end{align}
omitting pullback. Here, as before, the primed divisors are proper transforms of the unprimed. 

\begin{figure}[h] 
\begin{center}
\includegraphics{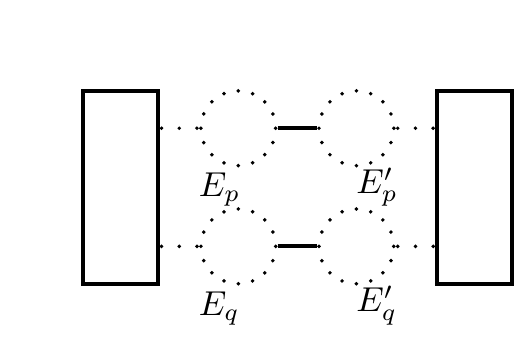}
\qquad
\includegraphics{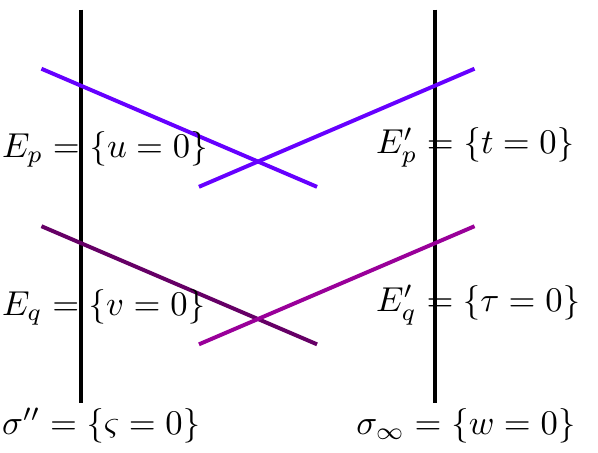}
\end{center}
\caption{\sl Blowing up two distinct instanton points give rise to double copy of E-string pairs, connected at different points of M9's.} \label{f:doubleE}
\end{figure}

This `bridge' $\{ E_q, E'_q \}$ over $q$ is parallel to the previous bridge $\{ E_p, E'_p \}$ over $p$, in the sense that none of the components has nonzero intersection with those. As a result, we have double copies of the above SCFT, as shown in Fig. \ref{f:doubleE}. 
For more blow-ups, as long as small instanton points are distinct, the resolution of them shall just be repetition.

\subsection{Coalescent cycles and induced $A_1$} \label{sec:inducedA1}

With the two intersection points $E_p \cap E_p'$ and $E_q \cap E_q'$ identified as 5-branes that are dual to M5-branes, we may make them coincident by bringing  together. For this, we may consider the difference between the corresponding cycles
\be \label{C1}
  C_1 \equiv E_p - E_q ,
\ee
which forms another divisor connecting the two 5-branes.
This cycle can be linearly equivalently expressed as $C_1 \sim E_q' - E_p'  $.
With Eq.(\ref{canonicalttF}), we can check 
\be \label{DuVal}
 K_{ \F_n^{(2)}} \cdot C_1=0.
\ee
It has intersection number
\be
 C_1 \cdot C_1 =  - 2,
\ee
forming the (minus of) Cartan matrix for the Lie algebra $A_1$.
The adjunction formula \cite{Shafa}
\be 
 K_{\F_n^{(2)}} \cdot C_1 + C_1\cdot C_1 = -2 + 2g
\ee
shows that the $C_1$ has zero genus $g=0$ and hence is a $\P^1$. 

This shows the McKay correspondence between the intersection numbers and the Cartan matrix of algebra. Blowing down the $C_1$ yields a surface singularity $\C^2/\Z_2$ of $A_1$. It is known that compactification of type IIB string theory on this singularity is mapped to {\em two} coincident M5-branes of $A_1$ type \cite{Strominger:1995ac,Witten:1995zh}, \cite{DelZotto:2014hpa,DelZotto:2014fia}. In fact, we have established even stronger correspondence between the weights of Lie algebras and the divisor class of E-string cycles. In the sequel, we shall extend it to $ADE$ algebras. 

As the cycle $C_1$ shrinks, two 5-branes are brought together. To do so, it is necessary to have linear equivalence relation $C_1 \sim E_p - E_q \sim 0$ by making the points $p$ and $q$ coincident. Note that this resolution does not modify the Calabi--Yau condition (\ref{CYcond}), because it is the resolution of du Val singularity, as in Eq.(\ref{DuVal}), so that the canonical class remains the same up to pullback \cite{GH}. Shrinking the cycle $C_1$ is done in the base without affecting the elliptic fiber, so there is no phase transition of small instanton type.

A D3-brane wrapped on the cycle $C_1$ is an M-string, dual to an M2-brane stretched between the above two M5-branes. From the construction (\ref{C1}), we may understand that two D3-branes wrapped on $E_p$ and $-E_q$ cycles can be continuously deformed into an M-string. As a matter of fact, this $A_1$  theory is different from that appears in six-dimensional ${\cal N} = (2,0)$ SCFT. First, here we have 7-branes dual to M9-branes, leaving the six-dimensional ${\cal N}  = (1,0)$ supersymmetry. So, the string is a charge source of the $(1,0)$ tensor multiplet. Second, as boundary condition from the 7-branes persists, the worldsheet gauge symmetry is not $U(q)$. This boundary condition also affects the projection on the gauge sector. We have seen in Sect. \ref{sec:Estring} that on E and \E-string worldsheets, we have respectively $O(q)$ and $USp(q)$ gauge theories. 
This means that six-dimensional ${\cal N} = (2,0)$ SCFT is recovered only locally in the close vicinity of the 5-branes; in this limit, 7-branes at $\sigma$ and $\sigma_\infty$ move away to infinity and we have no longer the boundary condition that breaks half of the supersymmetry \cite{Strathdee:1986jr}.

When we take $p$ and $q$ coincident, the gauge group is enhanced as $ O(q) \times USp(q) \to U(q)$. The corresponding vector and hypermultiplets are also enhanced as
\be \label{Uqenhancement} \begin{split}
 &\bf  \frac{q(q-1)}{2} + \frac{q(q+1)}{2} \to q^2\\
 &\bf  \frac{(q+2)(q-1)}{2} +1 + \frac{(q-2)(q+1)}{2} +1  \to q^2.
 \end{split}
\ee
The presence of the singlets of $O(q)$ and $ USp(q)$ is crucial in this enhancement structure. We shall verify this process later by surveying structure of the anomaly on the worldsheet. 

For later use, on each ends of the M-string of $A_1$ theory, we consider a  minimal but nontrivial dressing by two E-strings that is continuously deformable to heterotic string. As we shall see later, the total anomaly carried by this combination of constituent strings is equal to that of the heterotic string (after modding out missing charges from ejected instantons). We shall refer to this requirement as  the `SCFT block condition.' In fact, the 5-brane quiver gauge theory related to so manufactured minimal block is free of six-dimensional gauge anomalies \cite{Blum:1997fw,Blum:1997mm}, so it provides a basic building block involving $A_1$ theory in the global context. 

\begin{figure}[t]
\begin{center}
\includegraphics{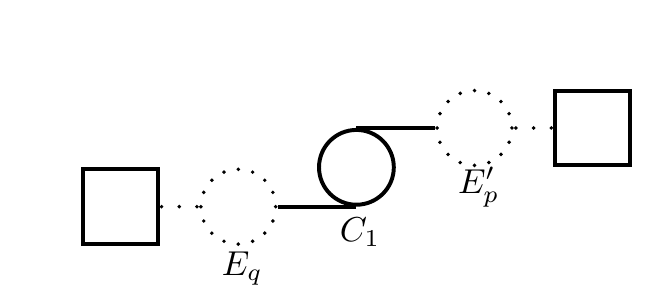} 
\qquad
\includegraphics{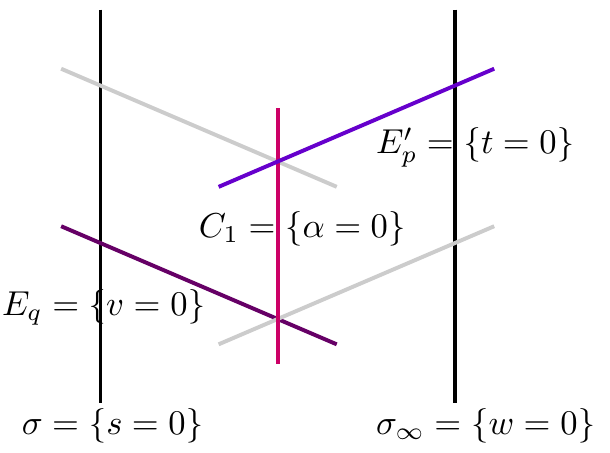}
\end{center}
\caption{\sl Solid (dotted) nodes denote the $\P^1$ cycles yielding M-strings (E-strings). Solid (dotted) lines denote the 5-(7-)branes. Although this shows the connection structure among three $\P^1$'s in the almost same way as Fig. \ref{f:doubleE}, the cycle corresponding to $C_1$ is induced as the difference between the other cycles, as in Eq.(\ref{C1}). \label{f:A1}}
\end{figure}

Our $A_1$ theory has the cycle $C_1 = E_p-E_q$ as defined in Eq.(\ref{C1}). The following divisor sum is linearly equivalent to $f$,
\be \label{A1MGSB}
 E_q + C_1 + E_p'  = E_q + (E_p - E_q) + (f - E_p) \sim f.
\ee
Therefore, we may additionally introduce E and E$'$-strings, respectively obtained by wrapping D3-branes on the cycles $E_q$ and $E_p'$, to have the combination equivalent to heterotic string. 
Among many possible linear equivalences, we have chosen the combination (\ref{A1MGSB}) because nothing but these
E and E$'$-strings are connected to $C_1$ intersecting once,
\be 
 E_q \cdot C_1 =  E_p' \cdot C_1= 1,
\ee 
while other strings from $E_p$ or $E_q'$ have negative intersections. 
The corresponding quiver diagram for this theory is derived from the $A_1$ Dynkin diagram, shown in Fig. \ref{f:A1}.

\subsection{More strings} \label{sec:morestrings}

So far, we have obtained E, M and heterotic strings by wrapping D3-branes on various combinations of two-cycles. Conversely, given a sum {$C$} of two-cycles in the sense of divisors, we may obtain various constituent strings. An interesting question is whether there are more building blocks than the previously identified ones. We are primarily interested in strings that can become tensionless, so we should be able to shrink the cycle $C$. Thus this $C$ should be rigid cycle and this condition singles out a rational, that is, genus zero curve. Recall that the genus of a curve $C$ is calculated by the Riemann--Roch theorem
\be \label{RRthm} 
 2 g(C) - 2 = K_B \cdot C + C \cdot C.
\ee
Since it depends on the canonical class $K_B$ of the whole base $B$, the {\em global} geometry of $B$ does matter. 
 
\subsubsection*{G-string, global case} 

We shall see that construction of $D$ and $E$ type ${\cal N} = (1,0)$ SCFTs requires a string from the cycle
\be \label{localDnroot}
 \tau = E_p + E_q.
\ee 
Although it has self-intersection $\tau \cdot \tau = - 2$, this cycle is not admissible because it has negative genus, using again Eqs. (\ref{canonicalttF}) and (\ref{RRthm})
\be \label{RRtheorem}
 2 g( \tau ) -2 = K_B \cdot \tau + \tau \cdot \tau  =  -1.
\ee

Consider next a modification of $\tau$ as
\be \label{Lcycle}
  C_G \equiv E_p + E_q  - f . 
\ee
This also forms a curve with self-intersection $(-2)$. Noting that $K_B \cdot f = -2$, we show that the $C_G$ has genus zero
$$ 
  g(C_G)   =\frac12 \left( K_B \cdot C_G + C_G \cdot C_G \right) +1 = 0.
$$
Wrapping a D3-brane on it yields another kind of string, that we may call a G-string. 
Since the $C_G$ is disjoint from $C_1$ considered above Eq.(\ref{C1}), $C_G \cdot C_1=0$, so we can independently blow down $C_G$ instead while leaving the individual component $E_p$ and $E_q'$ finite. This yields another nontrivial ${\cal N} = (1,0)$ SCFT of $A_1$ type having a tensionless G-string. Blowing down both $C_1$ and $C_G$ simultaneously means we also shrink the fiber $f$ itself, going to weakly coupled heterotic string. 

\begin{figure}[h]
\begin{center}
\includegraphics{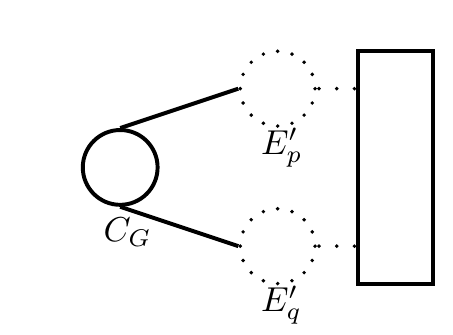}
\end{center}
\caption{\sl We can make a G-string by wrapping a D3-brane on the cycle $C_G$ in Eq.(\ref{Lcycle}). It is viewed as a linear combination of E and E$'$-strings, which can be attached to two E$'$-strings. This G-string will be needed in global construction of (1,0) $D$ and $E$ type SCFTs. \label{f:Gstring}}
\end{figure}

Viewed as a combination of E- and E$'$-string, 
$$
 C_G \sim f - E_p' - E_q'   \sim E_p - E_q' \sim E_q - E_p' 
$$
this G-string connects two different 5-branes. 
Among all possible combinations of these, only the $E_p'$ and $E_q'$ cycles have both intersection 1 with $C_G$ as $E_p' \cdot C_G = E_q' \cdot C_G=1$.
Using this, the SCFT block condition is satisfied as
$$E_q' + C_G+ E_p' = E_q' + (E_p - E_q') + (f - E_p) \sim f.$$
With the connection structure $E_p' \cdot \sigma_\infty = E_q' \cdot \sigma_\infty =1$, we can draw the Dynkin diagram as in Fig. \ref{f:Gstring}. As a result, although we have two E-strings attached from the right, the sum of cycles completes to the fiber $f$. 

\subsubsection*{Local case}

It is interesting to ask what happens in the local limit.
In the local description, we may construct the SCFT of $D_k$ type using the root of type in Eq.(\ref{localDnroot}).
The genus zero condition as dictated by the condition (\ref{RRtheorem}) can be evaded if we consider only local geometry because we have only part of the base $B' \subset B$ whose canonical class may have zero intersection $K_{B'} \cdot \tau =0$, to have 
$$
 2g(\tau) -2 = K_{B'} \cdot \tau + \tau \cdot \tau = -2. 
$$
Shrinking this cycle means $\tau \sim 0$, or
\be
 E_p \sim - E_q .
\ee
Recalling that $E_p$ is the cycle departing from the cycle $\sigma$, this reflection is possible with respect to the `left wall' $\sigma$. In other words, the local ${\cal N}= (1,0)$ SCFT of D-type should be in the vicinity of the wall.
  
\subsection{Conjugate strings}

We have defined \E-string, the conjugate to E-string, as the string obtained by wrapping a D3-brane on the negative divisor $-E_p$ to that of E-string, in Section \ref{sec:Estring}. Recall that both D3-branes are the BPS states preserving the same supersymmetries, with the orientation of E- and \E-strings in the remaining directions are also opposite.

This construction is generalizable to every composite string. A D3-brane wrapped on a two-cycle $C$ can have the same BPS state by wrapping another D3 on $-C$ with the opposite orientation of the remaining worldsheet. In forming cycles of non-Abelian structure, 
\be \label{variantcycles}
 C_i \sim E_{p} - E_{q} = (- E_{q}) - (- E_p) \sim  E_{q}' - E_{p}' = ( - E_p') - (E_q'),
\ee
and this exemplifies that various combination of strings are often equivalent. There is no natural preference of $E$ over $E'$. 

 Also, there is no natural preference between the cycle $C_i$ and the cycle $-C_i$. 
As such, we can make a SCFT of the same type using the conjugate strings from the cycles $-C_i$. However, the global structure is slightly different because there are different E-strings attached to them with positive intersection numbers. For the cycle $-C_i$, we now have $E_p \cdot (-C_i) = E_{p+1}' \cdot (-C_i)=1$. The anomaly free condition would be changed.

For example, we may define a conjugate G-string as a D3-brane wrapped on the cycle
\be
-C_G = E_p' + E_p' -f  \sim f- E_p - E_q.
\ee
Although it may not look effective, it is not irreducible and defined as difference, as in $C_G$. So we have no reason to neglect this possibility. The SCFT block condition connects two E-strings on the left,
 $$ E_p \cdot C_R = E_q \cdot C_R = 1,\quad C_R + E_p + E_q \sim f. $$
 \begin{figure}[h]
\begin{center}
\includegraphics{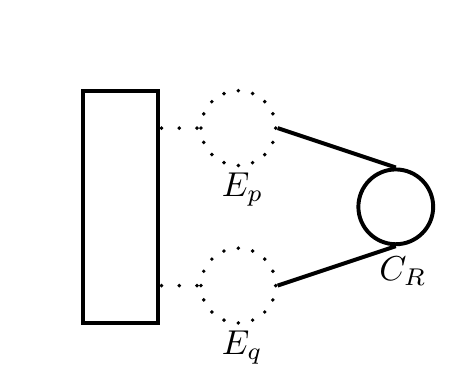}
\end{center}
\caption{\sl Conjugate-G-strings. A D3-brane wrapped on this cycle $C_R$ is the same BPS state as the D3-brane wrapped on the previous $C_G$-cycle. The global structure is different from the previous one because the two E-strings, instead of E$'$-strings can be attached to fulfill the SCFT block condition.
 \label{f:minusEstring}}
\end{figure}

\section{Non-Abelian Structure} \label{sec:nonabelian}

We may go on by blowing up more small instanton points and have as many exceptional divisors $E_i$. 
By making linear combinations, we may obtain various strings controlling 5-branes. We can construct $ADE$-type systems of arbitrary rank. Locally they  generate  six-dimensional SCFT's \cite{Strominger:1995ac,Witten:1995zh}. So far, generation of $ADE$ system using this reducible structure has not been discussed. We shall see that theories having $D$ and $E$ type structure need nontrivial completion in the global geometry. It is important that this rich structure is not easily caught in the M-theory or heterotic dual picture.
 
Consider again blow-ups at $k$ disjoint small instanton points $\{p_1, p_2, \dots, p_k\}$ in the base $ \F_n^{(k)} \to \F_n$.
We have exceptional divisors $\{ E_{p_1}, E_{p_2}, \dots E_{p_k} \}$, and the resulting proper transforms  $\sigma^{(k)} \sim \sigma - \sum E_{p_i}$. We also have $\{ E_{p_i}' \sim f - E_{p_i} \}$. The resulting canonical class is
\be
 K_{ \F_n^{(k)}} = -2 \sigma -(n+2) f + \sum_{i=1}^k E_{p_i} .
\ee
Again, with this, we have the modified Calabi--Yau condition (\ref{CYcond}). We have upper bound on the blow-up $k$ to be 24, which is the total number of instantons.

\subsection{$A_{k-1}$} \label{sec:Ak}

Generalizing the above discussion, we can construct theories of $A$-type. First take the combination,
\be
 C_i \equiv E_{p_i} - E_{p_{i+1}}, \quad i=1,\dots k-1,
\ee
to generate Lie algebra $A_{k-1}$. We have the intersections 
\begin{align}
& C_i \cdot C_i  = -2, \quad  K_B \cdot C_i = 0, \quad i=1,\dots k-1,  \\
 & C_i \cdot C_{i+1} = 1, \qquad \qquad \qquad \quad i=1,\dots k-2,
 \end{align}
while all others vanishing. Indeed, they have the same structure as the $A_{k-1}$ Dynkin diagram. Ordering of $E_{p_i}$ does not affect the connectivity structure, so we have natural definitions for $C_i$'s.

Shrinking all $C_i$'s, we obtain local geometry of $\C_2/\Z_k$ orbifold. In the M-theory side, the setup corresponds to a stack of $k$ parallel M5-branes. Then, locally, we have six-dimensional $A_{k-1}$ theory which still has ${\cal N}= (1,0)$ supersymmetry. 

\begin{figure}[h]
\begin{center}
\includegraphics{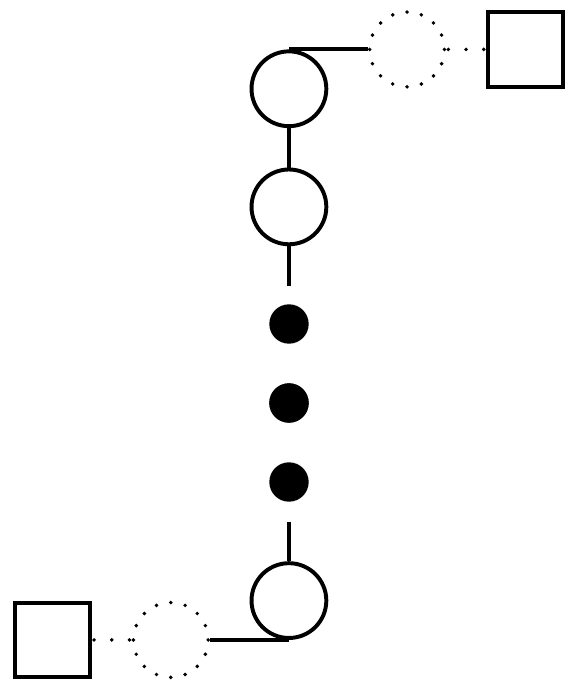}
\end{center} 
\caption{\sl Typical quiver diagram for $A_{k-1}$ type. The left E-string and the right E-string should be at the bottom and the right E-string at the top.}\label{f:Aquiver}
\end{figure}

We seek the SCFT block condition involving this $A_{k-1}$ theory, as we did  in the last section. Note the relation
\be \label{heteroticstructure}
\sum_{i=1}^{k-1} a_i^\vee C_i = E_{p_1} - E_{p_k} \sim -  E_{p_1}' - E_{p_k} + f , 
\ee
where $a_i^\vee=1$, for all $i$, are Dynkin labels for $A_{k-1}$. To complete this, we should wrap $a_i^\vee$ D3-branes on each cycle $C_i$. Also we have to attach E$'$-string departing from $C_1$ and E-string ending on $C_k$.
$$ E_{p_{k}} + \sum_{k=1}^{k-1} a_i^\vee  C_i + E_{p_1}' \sim f. $$
Each of them intersects the $A_{k-1}$ once $E_{p_1}' \cdot C_{1} = E_{p_k} \cdot C_{k-1}=1$. About the $E_8$ loci, the proper transform of $\sigma^{(k)}$ and $\sigma_\infty$, which we respectively name $\sigma^{(k)}$ and $\sigma_\infty$, have intersections $\sigma^{(k)} \cdot E_{p_i}=1=E_{p_i}' \cdot \sigma_\infty$ for every $p_i$ above. The corresponding quiver diagram is shown in Fig. \ref{f:Aquiver}.

\subsection{Affine extension $\widehat A_{k-1}$}

Extending the above relation between the orthonormal basis and $E_{p_i}$ cycles for the simple Lie algebras $A_{k-1}$, we may attempt to extended to $\widehat A_{k-1}$ theory by introducing a divisor corresponding to the extended root
\be
C \equiv E_{p_k} - E_{p_1}.
\ee
However, this cycle is not independent because 
$$ C + \sum_{i=1}^{k-1} C_i =0. $$

With another cycle $\nu$ in the base $B$, we may form a new cycle
\be \label{nucycle}
 C_0 \equiv E_{p_k} - E_{p_1} + \nu,
\ee
having desired intersection structure 
\be
 C_0 \cdot C_0=-2, \quad C_{k-1} \cdot C_0 = 1.
\ee
This requires the intersection structure on $\nu$
\be \label{nullcycle}
 \nu \cdot E_{p_k} = \nu \cdot E_{p_1} = 0, \quad \nu \cdot \nu = 0.
\ee
We find this desirable, as we can naturally relate the cycle $\nu$ with the imaginary root of the $\widehat A_k$ having zero norm.
We also need that the genus of $C_0$ should be zero
\be
 -2 + 2 g(C_0) = K_B \cdot C_0 + C_0 \cdot C_0 = -2,
\ee
yielding
\be
 K_B \cdot \nu =0.
 \ee
 This implies that the $\nu$ is a genus one curve or a real torus
\be
  \quad K_B \cdot \nu + \nu \cdot \nu = 0 = -2 + 2g(\nu).
\ee

For the obvious reason, we may call the cycle $\nu$ null cycle. This genus one curve contains non-contractable circle ${\mathbb S}^1$. This has the same structure of the imaginary root of the affine Lie algebra \cite{Fuchs}. The strings are classified by wrapping number of $C_i$'s and $\nu$ which form the weight under the affine Lie algebra. 

\begin{figure}[h]
\begin{center}
\includegraphics{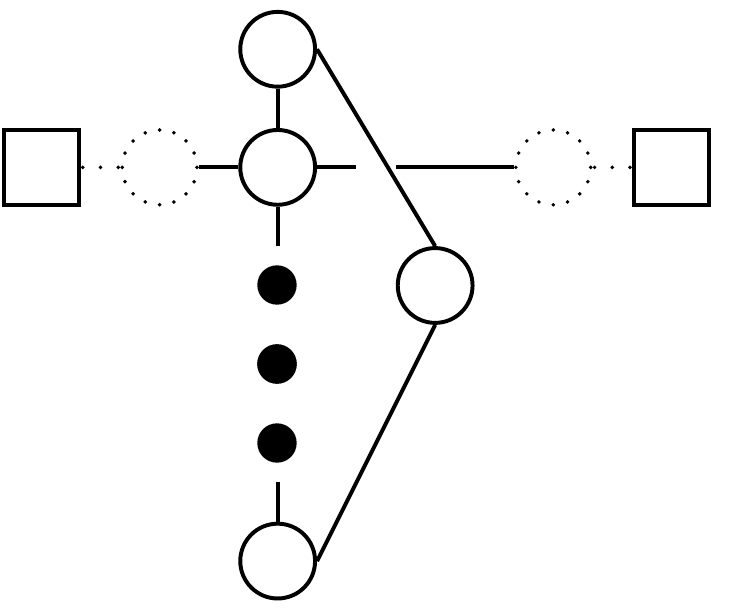}
\end{center} 
\caption{\sl Typical quiver diagram for $\widehat A_{k-1}$ theory. To have a nontrivial cycle, the closed quiver should be on the genus one cycle, or the M5-branes should be aligned along the corresponding non-contractable circle. As long as a pair of E-strings is formed, we can attach it to any node. We can decouple such E-strings in a local limit to have six-dimensional ${\cal N} = (2,0)$  SCFT of $\widehat A_{k-1}$ type that is free of gauge anomalies.}\label{f:hatAquiver}
\end{figure}

Let us consider SCFT block condition. The relevant relation is
\be \label{Akstructure}
 \sum_{I=0}^{k-1} a_I^\vee C_I = \nu,\quad a_0^\vee \equiv 1,
\ee
in which we have no contribution from $f$.
Since $E_{p_j} \cdot C_{j-1} = E_{p_j}' \cdot C_{j} = 1$ is the only possibility for intersection\footnote{It is understood that $C_{-1}  \equiv C_{k-1}$.}, a pair of E-strings departing and ending at the same 5-brane are necessary to complete the trip. They may be attached to any node because of cyclic symmetry of $\widehat A_{k-1}$ root system. This situation is drawn in Fig. \ref{f:hatAquiver}. 

In other words, we have two superimposed theories in six dimension. One is a heterotic string and the other is a locally ${\cal N} = (2,0)$ SCFT of $\widehat A_{k-1}$ type. The relation (\ref{Akstructure}), in contrast to Eq.(\ref{heteroticstructure}), shows that the heterotic string now is a spectator, so that we can decouple them in the decoupling limit that we will consider later in Sect. \ref{sec:decoupling}.

The resulting $\widehat A_{k-1}$ theory is gauge anomaly free on its own \cite{Blum:1997fw,Blum:1997mm}. Thus we can also extend the SCFT block condition to the combination of strings that is not only reduced to heterotic string, but also to to a string from any null cycle as in 
Eq.(\ref{Akstructure}). However, we will see later that, to guarantee the gravitational anomaly cancellation we should place this theory to global construction along the heterotic string from the cycle $f$.

\subsection{$D_k$}
Adapting to the root structure, we may consider the following collection of E-string cycles having the same connection structure as the  $D_k$ Dynkin diagram,
\begin{align}
& C_i  \equiv E_{p_i} - E_{p_{i+1}}, \quad i=1,\dots k-1,\\
& C_{k}  \equiv E_{p_{k-1}} + E_{p_k}  -f . \label{LinDk}
\end{align}
In the last line, we need to introduce G-string as in Sect. \ref{sec:morestrings} due to the global geometry. A naive cycle  $E_{p_{k-1}} + E_{p_k}$ cannot be genus zero without adding the cycle $f$. It can be also expressed as $C_k \sim E_{p_k} - E'_{p_{k-1}} \sim f-E_{p_k}' - E_{p_{k-1}}' $.

They span the root lattice, which can be checked by the intersections
\be \label{Dkstruct} \begin{split}
 C_i \cdot C_{i+1} &= 1,\quad  i=1,\dots k-1, \\
 C_k \cdot C_{k-2} &= 1,
 \end{split}
\ee
with all others vanishing.

\begin{figure}[t]
\begin{center}
\includegraphics{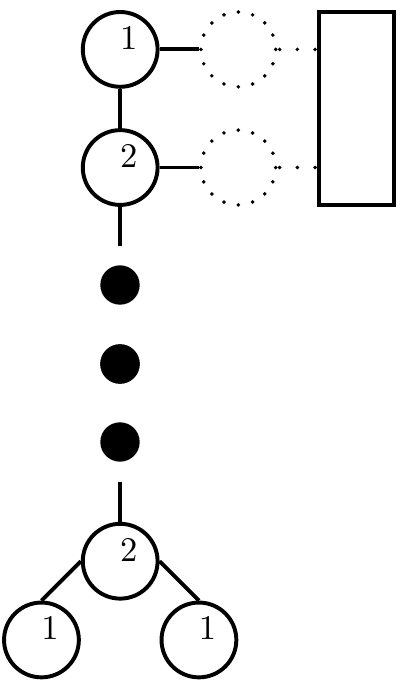} 
\end{center}
\caption{\sl Typical `creeper' quiver diagram for $D_k$ theory. We need G-string considered in Sect. \ref{sec:morestrings}. The numbers in the nodes denote the Dynkin labels for $D_k$. We have as many M-strings to have anomaly free theory.}
\label{f:Dk}
\end{figure}

Blowing down all the cycles $C_i,i=1,\dots k$ gives rise to $\C^2/{\mathbb D}_{k-2}$ singularity, where ${\mathbb D}_{k-2}$ is a binary dihedral group of order $(k-2)$. The corresponding quiver diagram is shown in Fig. \ref{f:Dk}.

We seek the SCFT block condition for consistency. One might consider wrapping one D3-brane on each cycle $C_i$,  but 
we have, for generic $k$,
\be \begin{split}
 \sum_{i=1}^k a_i^\vee C_i = & C_1 + 2 C_2 + 2 C_3 + \dots  + 2 C_{k-2} + C_{k-1} + C_k \\
  &=  E_{p_1} + E_{p_2} - f \sim - E_{p_1}' - E_{p_2}' +f 
 \end{split}
\ee
where $a_i^\vee$ are Dynkin labels for $D_k$. Therefore, the SCFT block condition becomes
\be
 E_{p_1}'+  E_{p_2}'+  \sum_{i=1}^k a_i^\vee C_i  =  f.
\ee
In particular, we need $a_i^\vee$ wrapping for each D3-brane on each $C_i$ cycle.  The only possible combinations of E-strings as wrapped D3-branes are $E_{p_1}'$ and $E_{p_2}'$. Every string is obtained by D3-brane wrapped on the cycles $\{ E_{p_1}', C_i, E_{p_2}'\}$. Indeed, they have positive intersections
\be
 C_1 \cdot E_{p_1}' = C_2 \cdot E_{p_2}'= 1.
\ee
In other words, in the Dynkin diagram, we can attach two E$'$-strings on the first two nodes.

In this global description admitting heterotic dual, the only way to obtain $D$-type SCFT is to make use of G-string. It is because of the structure of the standard root $C_k$ in Eq.(\ref{LinDk}).  For the crepant resolution condition $K_B \cdot C_i = 0$, required for well-defined shrinking to obtain $D_k$ singularity, we need the curve to have self-intersection $(-2)$ and genus zero. This is to be distinguished from the local resolution where we do not have the contribution from $f$. Thus, in this case, the $A_3$ theory is not the same as the $D_3$. The $A_3$ theory makes use of four 2-cycles $E_{p_i},i=1,2,3,4,$ from as many detached small instantons, while the $D_3$ theory uses only three $E_{p_i},i=1,2,3$'s and $f$. 

We can affinize $D_k$ theory as in the previous section. With the null cycle of genus one $\nu$ as in Eq.(\ref{nullcycle}), we can introduce the cycle corresponding to the extended root, 
\be \label{C0}
 C_{0} \equiv - E_{p_{1}} - E_{p_{2}}  + f + \nu,\quad a_0^\vee \equiv 1,
\ee
forming the affine $\widehat D_k$ theory. We can again verify that $C_0$ has genus zero, for which the existence of the $f$ component is crucial. The sum, now including $C_0$,
\be 
 \sum_{I=0}^k a_I^\vee C_I =  \nu
\ee
is independent of $f$. That is, the cycle $f$ decouples as in the case of the affine $\widehat A_{k-1}$. Therefore we can isolate ${\cal N} =(2,0)$ $\widehat D_k$ SCFT from the heterotic string theory in the decoupling limit. Similar extensions are possible for other cases that we shall discuss now.

\subsection{$E_8$ and others}

Likewise, we can form $E_6,E_7, E_8$ lattices using E-string cycles, yielding 5-brane configurations sharing the same names. For example, blowing up eight instanton points, we have $E_{p_i},i=1,\dots8$. With the following two-cycles $C_i$, we may form resolved $E_8$ singularity in the base $B$: 
\begin{align}
 C_i &\equiv E_{p_{i+1}} - E_{p_{i+2}}, \quad i=1,\dots 6.\\
 C_7 &\equiv \frac12 E_1 - \frac{1}{2} \sum_{i=2}^7 E_{p_i} + \frac12 E_8 +f, \label{obliquestring}\\
 C_8 &\equiv E_{p_7} + E_{p_8} -f .
\end{align}
For constructing $C_7$, we should admit half two-cycles as components. The existence of such generator should be justified. A known consistency condition is that the Dirac quantization for the string charges should be integral \cite{Seiberg:2011dr}. This is translated to the condition that the whole lattice still have integral products, understood as intersection numbers. Another condition is self-duality of the lattice, which is automatically satisfied by Poincar\'e duality in F-theory construction \cite{Seiberg:2011dr}. As in the case of $D_k$, we need additional contribution of $f$ in $C_7$ and $C_8$ to make them genus zero in the global geometry.

\begin{figure}[t]
\begin{center}
\includegraphics{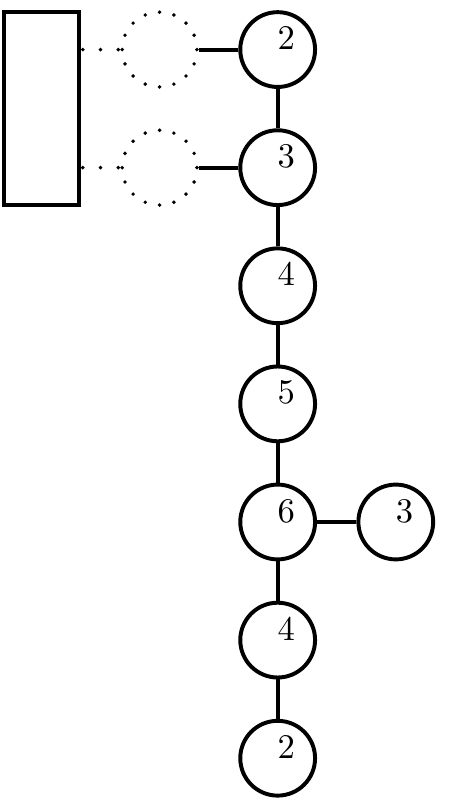}
\end{center}
\caption{\sl Quiver diagram for $E_8$ theory. We also need G-string in the global construction.} The same conventions as above. \label{f:e8dynkin}
\end{figure}

With the Dynkin labels $a_i^\vee$ of $E_8$, we may have linear relation
\be \begin{split}
 \sum_{i=1}^8 a_i^\vee C_i  &=  2 C_1 + 3 C_2 + 4 C_3 + 5 C_4 + 6 C_5 + 4 C_6 + 2 C_7 + 3 C_8  \\
 &=   - E_{p_1}  - E_{p_2} + f.
\end{split}
\ee
That is, blowing down all the cycles $C_i$ gives rise to a singularity $\C^2/{\mathbb I}$ where $\mathbb I$ is a binary icosahedral group.
The SCFT block condition can be taken by connecting the two E-strings from the left $E_8$ to $C_1$ and $C_2$, and by wrapping D3-branes on $C_i$ by $a^\vee_i$ times
\be
E_{p_1}+ E_{p_2} +  \sum_{i=1}^8 a_i^\vee C_i \sim f.
\ee
We have the resulting quiver diagram (or E-strings connected to the left) in Fig. \ref{f:e8dynkin}.

Other lattices $E_6$ and $E_7$ can be similarly constructed as subgroups of $E_8$. We can have only simply laced algebra \cite{Henningson:2004dh,Shimizu:2016lbw}. Other type of algebras $B_k, C_k, F_4, G_2$ and their affinization involve the roots whose self-intersection matrix is asymmetric. Although we may obtain such by linear combinations of $E_i$'s, it fails to be $\P^1$, as adjunction formula shows. Note that we have classified all possible six-dimensional SCFT's of M5-branes of $ADE$ type, simple and affine.

\subsection{Little strings and decoupling limit} \label{sec:decoupling}

Here we pause to understand the gravity decoupling (local) limit and the little string (rigid) limit, respectively. It is recently recognized \cite{Bhardwaj:2015xxa, Hohenegger:2015btj, Bhardwaj:2015oru, Hohenegger:2016eqy, HIRtoappear} that the F-theory construction of M5-branes provides clean description on little string theories.
Strictly speaking, there are two kinds of little strings. One is heterotic string from a D3-brane wrapped on the cycle $f$. Another is little string from a D3-brane wrapped on any local genus one cycle $\nu$. Both cycles have zero self-intersections.

Recall the duality between F-theory compactified on elliptic K3 surface over the $\P^1$ base $f$ and heterotic string on torus $T$.
A D3-brane wrapped on $f$ in the F-theory side is mapped to heterotic string. Taking Hodge dual, a D3-brane wrapped on $\sigma$ gives another string in the normal direction, which should be an NS5-brane wrapped on $T$ and $\sigma$ of the heteroic side. Their tensions are related as
\be \label{tensions}
 T_{\rm h} = T_{\rm D3} \vol f \qquad \mbox{and} \qquad T_{\rm NS5} \vol T = T_{\rm D3}. 
\ee
The volume of the fibers does not change under the deformation or resolution.
Letting the heterotic string tension be $T_{\rm h} = 1/(2 \pi \ell_{\rm h}^2)$ and  type II string tension be $1/(2\pi \ell_{\rm s}^2)$, these relations become
$$ 
\frac{1}{2 \pi \ell_{\rm h}^2} =  \frac{2 \pi}{(2 \pi \ell_{\rm s})^4 g_{\rm  B}}  \vol f 
 \qquad \mbox{and} \qquad \frac{2\pi}{(2 \pi \ell_{\rm h})^6 g_{\rm h}^2}  \vol T=  \frac{2 \pi}{(2 \pi \ell_{\rm s})^4 g_{\rm  B}}, $$
respectively, with IIB coupling $g_{\rm B}$ and heterotic coupling $g_{\rm h}$.
Combining, we obtain the ten-dimensional heterotic string coupling
\be \label{heteroticcoupling}
 g_{\rm h}^2= \frac{\vol T \vol f}{(2 \pi \ell_{\rm h})^4}
\ee
in terms of the volumes measured in the string unit.

Let us focus on the heterotic string side. The ten-dimensional gravitational coupling is
\be
 2 \pi \kappa_{\rm h}^2 =  (2\pi \ell_{\rm h})^8 g_{\rm h}^2  = (2\pi \ell_{\rm h})^4 \vol T \vol f.
\ee
Dimensional reduction shows that the eight-dimensional gravity coupling is essentially the volume of cycle $f$ from which heterotic string arises \cite{Vafa:1996xn,Donagi:2008ca,Beasley:2008dc}
\be
2 \pi \kappa_{\rm 8D}^2 = (2\pi \ell_{\rm h})^4 \vol f = \frac12 g_{\rm YM,8}^2 \ell_{\rm h}^2.
\ee

Further reduction gives six-dimensional gravitational coupling
$$ 
 2 \pi \kappa_{\rm 6D}^2 =(2\pi \ell_{\rm h})^4  \frac{\vol f }{ \vol \sigma } . 
$$
We may decouple gravity $\kappa_{\rm 6D} \to 0$ by decompactifying the base $\vol \sigma \to \infty$. Still, we can keep the volume of the fiber $f$ finite. As a six-dimensional physics, we can see only localized SCFT on the 5-branes. 
Also we can turn off string interaction $g_{\rm h} \to 0$ in Eq.(\ref{heteroticcoupling}) by taking $\vol T \to 0$. In effect, we have decoupled local SCFT from the bulk physics. 
This is characterized by the scale 
$$ M^2 \sim \frac{1}{\vol f}. $$

We can decouple eight-dimensional gravity by sending the entire volume to infinite $\vol f \to \infty, \vol \sigma \to \infty, (\vol f / \vol \sigma) \to 0$.  We call it decoupling limit. Now, each stack of M5-branes in the bulk have no interaction, so we may have collection of local six-dimensional ${\cal N} = (2,0)$ SCFTs.

Now, we consider the SCFT limit, focusing on the type IIB side.
The gravitational coupling is
$$
 2 \pi \kappa_{\rm B}^2 = (2 \pi \ell_s)^8 g_{\rm B}^2. 
$$
We have 7-branes wrapped on $E_{p_i}$.
Fixing everything here, we have eight-dimensional coupling for the theory on $E_{p_i}$
\be
 2 \pi \kappa_{\rm B,8D}^2 = \frac{(2 \pi \ell_s)^8 g_{\rm B}^2}{\vol E_{p_i}} = g_{\rm YM,8}^2 \ell_s^2.
\ee
That is, the coupling $g_{\rm YM,8}^2$ of the gauge theory on the eight-dimensional worldvolume is inversely proportional to the volume of the cycle, which is parametrized by the scalar field of tensor multiplet. This becomes the coefficients of anomaly polynomials \cite{Seiberg:1996qx,Morrison:1996na}. This means that we have superconformal limit when we have no dimensionful parameter, that is, in the shrinking limit of the cycle $E_p$, in which gauge coupling diverges.

Various two-cycles that we have obtained so far are related to tensor multiplets of six-dimensional ${\cal N} = (1,0)$ supersymmetry. Among them, there is a special tensor multiplet containing the dilaton of the heterotic string as a scalar component. This is proportional to the volume of the fiber $f$ of the Hirzebruch surface $\F_n$. The same holds true for other derived tensor multiplets from blow-up, whose scalar components proportional to the volumes of $C_i$ give rise to dilaton of the corresponding 5-brane worldvolume gauge theories.
We have also little string theory obtained by a D3-brane wrapped on $\nu$ in Eq.(\ref{nucycle}). The tension $T$ is again proportional to the volume $\vol \nu$. Thus $T$ is finite if $\vol \nu$ is finite.


\section{Coalescent Small Instantons}
We may also consider more than one small instantons shrunk at the same point, giving severer singularity. The resulting point should be described as many coincident 5-branes. They can be also extracted from the 7-brane on $\sigma$ by a series of resolutions, revealing non-Abelian structure. { This process has been well-understood in the previous studies  \cite{Aspinwall:1997ye,Heckman:2013pva,DelZotto:2014hpa}. Now, the natural question here is whether these results are compatible with the new processes discussed in Sections \ref{sec:Mstrings} and \ref{sec:nonabelian}, where we had distinct instantons at different points. We will also see that the full resolution of the coalescent small instantons is more involved and requires extra physics in the bulk.} 

Let $k$ small instantons be coalescent at the point $z'=0$ in one $E_8$ at $z=0$. It is reflected in the Weierstrass equation as
\be \label{tunedm5}
 y^2 = x^3 + \left( f_8 z^4 w^4 \right) x + \left( g_{12-n-k} z^5 z^{\prime k}  w^7 + g_{12} z^6 w^6 + g_{12+n} z^7 w^5 \right),
\ee
having the discriminant
$$
 \Delta = z^{10} w^{10} \left( 4 f_8^3 z^2 w^2 + 27 g_{12-n-k}^2  z^{\prime 2k} w^4 + 54 g_{12-n-k} g_{12} z z^{\prime k} w^3 + \dots \right) . $$    

{\em Locally} around $z \simeq 0$, the curve (\ref{tunedm5}) is reduced to
\be
 y^2 \simeq x^3 + z^{\prime k} z^5.
\ee
The possible approximate singularities $\C^2/ \Gamma_G$ at $x=y=z'=0$ for various $k=1,2,3,4,5$ can be respectively $\rm II, IV, I_0^*, IV^*, II^*$, as displayed in Table \ref{t:smallinsts}. This singularity is not globally extended along $z'=0$, but only reflects local shape of the divisor $D'_{\text{inst}}$ in the vicinity of $z=0$, as shown in Figs. \ref{f:blow-up}, and \ref{f:twoEs} for the case of $k=1$. Having larger $k$ means deforming the shape of the $D'_{\text{inst}}$.

Therefore, this situation should be contrasted against the one where $z'=0$ is a {\em globally} well-defined discriminant locus over the entire base $B$. To do so, in addition to the $g_{12-n}$ above, we may also tune all the coefficient including $f_8, g_{12}$ and $g_{12+n}$ to have a factor $z'$. The discriminant divisor in Eq.(\ref{disclocus}) now has the form
$$
 D = 10 \sigma + 10 \sigma_{\infty} + 2k f + D',
$$
where $D'=4 \sigma + (2n+24-2k)f$ \cite{Aspinwall:1997ye}. Here, having larger $k$ means putting more 7-branes on $z'=0$. This can be more clearly seen in the corresponding Weierstrass equation, as presented in the appendix A.

\subsection{Regular resolution} \label{s:regularres}

\begin{figure}[h]
\begin{center}
\includegraphics{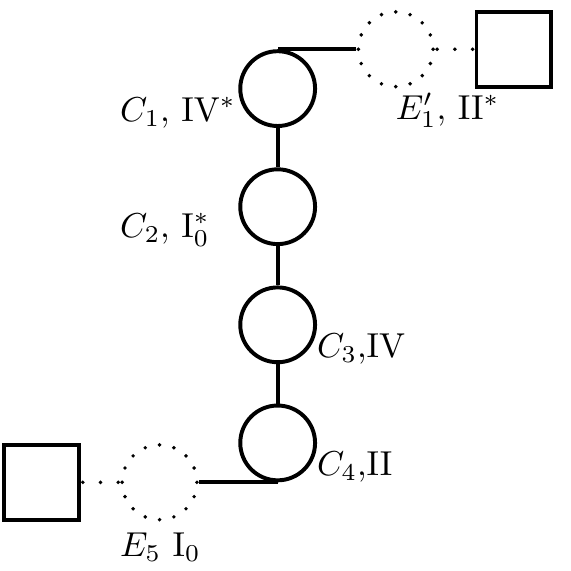}
\end{center}
\caption{\sl Partial resolution of local singularity with $k=5$, of a form $y^2 = x^3 + (z')^5 z^5.$ Although the supported singularities are different, the base geometry is identical to the previous $A_4$ case. Compare with Fig. \ref{f:Aquiver}. \label{fig:A4res}}
\end{figure}

To be concrete, we shall analyze the most singular case, $k=5$. Locally around $z \simeq 0$, we have $\I\I^*$ singularity at $z'=0$. As discussed in \cite{Aspinwall:1997ye,Gr,Mi}, we may iteratively blow up in the base $z=z'=0$
\begin{equation} \label{regreschain} \begin{split}
\sigma &\sim \sigma^{(1)}+ E_1 , \quad  f \sim  E_1 +E'_1, \\
\sigma^{(1)} &\sim \sigma^{(2)}+ E_2 , \quad E_1 \sim  E_2  + C_1, \\ 
\sigma^{(2)} &\sim \sigma^{(3)}+ E_3 , \quad E_2 \sim  E_3 + C_2, \\
\sigma^{(3)} &\sim \sigma^{(4)}+ E_4 , \quad E_3 \sim  E_4 + C_3,  \\
\sigma^{(4)} &\sim \sigma^{(5)}+ E_5 , \quad E_4 \sim  E_5 + C_4.
\end{split}
\end{equation}
The resolution ends here because the small instanton point is no more singular beyond Kodaira.
We arrived at the smooth fiber $\I_0$ at $E_5$.
Here, $E_i,i=1,\dots 5$ are exceptional divisors with $E_i \cdot E_i = -1$ and $C_i,i=1,\dots 4$ are the proper transforms, so that
$$
K_{\F^{(5)}} = -2 \sigma - (n+2) f + \sum_{i=1}^5 E_i .
$$
As we have seen in Eq.(\ref{SigmaprimeIntersec}), blow-up and proper transform modify the self-intersection number of the divisor containing the point. For example, in Eq.(\ref{regreschain}) we obtain the proper transform $C_i$ from $E_i$, 
$$ E_i \cdot E_i=-1 \quad  \Longrightarrow \quad C_i \cdot C_i = (E_i - E_{i+1} ) \cdot (E_i - E_{i+1} )  =-2.  $$
Therefore, the resulting set of divisor $C_i$ has the intersection structure of the $A_4$, as shown in Fig. \ref{fig:A4res}. Summarizing
\begin{align}
 E_1^{\prime } \cdot E_1^{\prime } &= {E_5} \cdot E_5 = -1, \\
 C_i \cdot C_i &= -2,  \quad i=1,2,3,4,\\
  C_i \cdot C_{i+1} &= 1, \quad i=1,2,3,\\
  E_1'  \cdot C_1 &= E_5 \cdot C_4 = 1,
\end{align}
and others vanish. For later reference, we may call this regular resolution chain.

Remarkably, this turns out to be {\em same resolution} as that of the previous $A_4$ theory considered in Section \ref{sec:Ak}, which is constructed from the five distinct small instantons. Although, in the end, we do not have $E_i,i=1,2,3,4$ and $E_j',j=2,3,4,5$, we may define them as linear combinations of the other $E_1',E_5,C_i$'s. We can see that these cycles are not discriminant loci, so there are no supported 7-brane gauge group in the global limit. 
So far, the processes commute as follows:
\vskip0.5cm
\begin{tcolorbox}
$$
\begin{CD}
    \text{distinct small instantons} @>{\text{tensor}}>>  \text{distinct 5-branes}\\
@VV{\text{Higgs}}V @VV{\text{Higgs}}V\\
\text{coincident small instantons} @>{\text{tensor}}>> \text{coincident 5-branes}.
\end{CD}
$$
\vskip0.2cm
\end{tcolorbox}
\vskip0.2cm
\noindent Here, the moduli space is traveled along the indicated branch. As in the four-dimensional counterpart involving vectors and hypermultiplets, the moduli spaces of tensor branch ${\cal M}_T$ and Higgs branch ${\cal M}_H$ are disconnected
$$ 
{\cal M}_T \times {\cal M}_H.
$$

Locally, around $z \simeq 0$, the divisors $E_5,C_4,C_3,C_2,C_1, E_1'$ support $\I_0,{\rm II}, {\rm IV}, \I_0^*,{\rm IV}^*,\IIstar$, respectively, as shown in Figure \ref{fig:A4res}.  We extracted five small instantons from the $E_8$ at $\sigma$ and now have as many 5-branes at the intersections. However in the global geometry there is no supported gauge theories on those divisors, because they are not discriminant loci.

\subsection{Special resolution}

The regular resolutions in the previous section have completely resolved the singularity at the ``boundary'' $z=0$. We had natural interpretation of extracting M5-branes seen in the M-theory dual side. However, the resolution may not look complete in the sense that collision between the $C_1,E_1'$ pair and $C_1,C_2$ are still singular beyond the Kodaira classification. Such intersections are now in the bulk, and have nothing to do with small instantons in the $E_8$. We may resolve first the intersection between $C_1$ and $E_1'$ as follows
\begin{equation}  \begin{split}
 C_1 &\sim  C_{-1} + C_1' , \quad E_1' \sim C_{1}' + B_2,  \\
 C_1'& \sim C_{-2}+ C_2',  \quad B_2 \sim C_{2}' + B_3,  \\
 C_2' &\sim C_{-3} +C_3',   \quad B_3 \sim C_{3}' + B_4, \\
 C_3' &\sim C_{-4}  + C_4', \quad B_4 \sim C_{4}' + B_5.
\end{split} \end{equation}
Here, $C_i'$ are exceptional divisors of self-intersection $(-1)$. 

\begin{figure}[t]
\begin{center}
\includegraphics{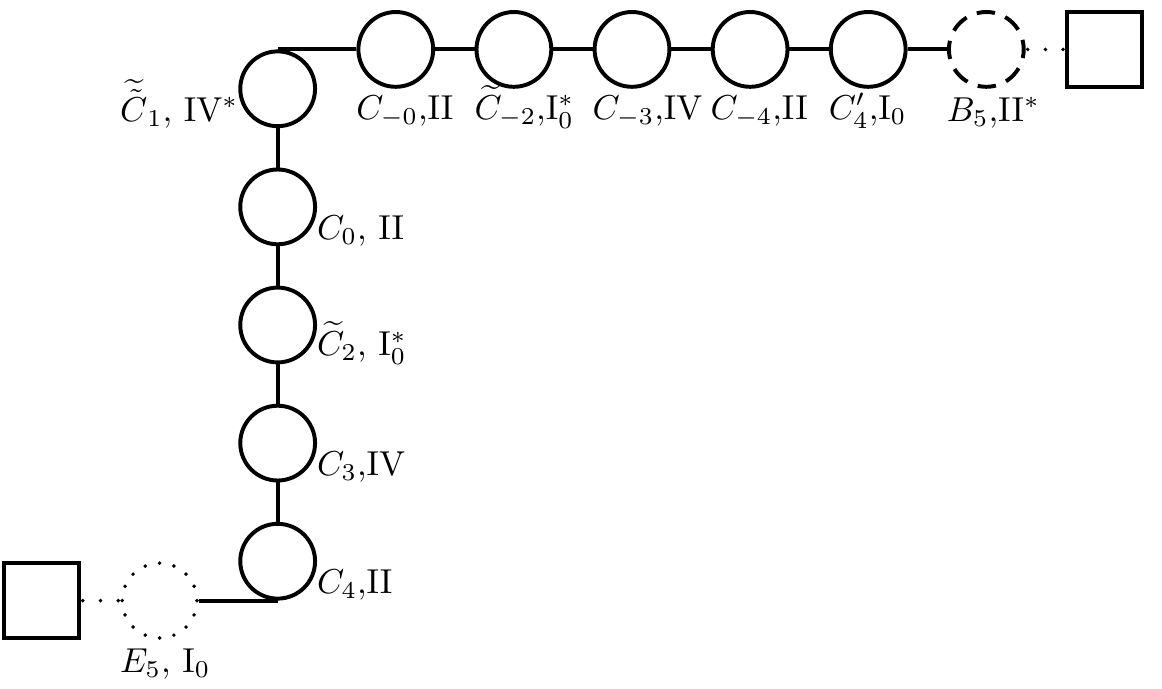}
\end{center}
\caption{\sl Full resolution of the $k=5$ case. The local supporting gauge group, seen from $z\simeq 0$ is determined by self-intersection except the rightmost node $B_5$. Also the singularity on $B_5$ and $\sigma_\infty$ does not collide to enhance the singularity.  \label{fig:FullresE8}}
\end{figure}

Note that the proper transform of the cycle $C_1$ is the cycle $C_{-1}$. Still, the intersection between local singularity $\rm IV^*$ on $C_{-1}$ and $\I_0^*$ on $C_2$ is too singular beyond the Kodaira classification, so that we resolve it as
$$
 C_{-1} \sim \widetilde C_1 + C_0, \quad  C_2 \sim  \widetilde C_2 + C_0,  
 $$
with the exceptional divisor $C_0$. We have the same singularitiy also at $C_{-1},C_{-2}$ that can be resolved as
$$
  \widetilde C_{1}  \sim \widetilde{\tilde C}_{1}+ C_{-0}  , \quad  C_{-2} \sim  C_{-0}+ \widetilde C_{-2},
$$
with the exceptional divisor $C_{-0}$.
This completes the resolution. The resulting canonical class is
\be
  K_B = -2 \sigma - (n+2) f + \sum_{i=1}^5 E_i  + \sum_{k=1}^4 C_k' + C_0 + C_{-0}.
 \ee
Again, all the processes are commutative.
The resulting cycles and the supported singularities are respectively shown in Fig. \ref{fig:FullresE8}. Remember that, although the rightmost node $B_5$ intersects $\sigma_\infty$ once, the vanishing local equation $(B_5)=0$ does not enhance the gauge symmetry $E_8$ located at $\sigma_\infty$, that is, it does not make $w^5 z^7$ term in Eq.(\ref{tunedm5}) vanish. In Appendix B, we provide a simpler example. We may call this as special resolution chain. We have intersection numbers
\be
  \widetilde{\tilde C}_{1} \cdot  \widetilde{\tilde C}_{1}  =B_5\cdot B_5= -5, \quad  {\widetilde C_2} \cdot {\widetilde C_2} =  {\widetilde C_{-2}} \cdot {\widetilde C_{-2}}  = -3,
\ee
while others remain the same. The result agrees with local analysis using maximal Higgsible cluster \cite{Morrison:2012np,DelZotto:2014hpa}, except the $B_5$ which does not belong to this.

Since all the resolution is done in the bulk and did not touch $\sigma$, we have not extracted small instanton of $E_8$. In fact, the special resolution cannot be done without ruining the Calabi--Yau condition (\ref{CYcond}) because the scaling cannot be done globally. If small instantonic 7-branes have globally well-defined coordinates, we are able to resolve it.

\subsection{Other cases}
Extending the discussion to the rest of cases is straightforward, since the ones with $k \le 4$ arise as an intermediate step of the previous analysis. For $k=4$, we have local ${\rm IV}^*$ at $z'=0$. We may follow the regular resolution chain (\ref{regreschain}) from the first to the fourth. Then, we have special resolution at the intersection between $ \I_0^* $ and ${\rm IV}^*$ singularities. It ends up with 
\be
 \IIstar - \I_0 - {\rm II} - {\rm IV} - \I_0^* - {\rm II} - {\rm IV}^*
\ee
For smaller $k$, we have only regular resolution chain (\ref{regreschain}) from the first line to the $k$'th line. For $k=3$, we have local $\I_0^*$
\be
 \IIstar - \I_0 - {\rm II} - {\rm IV} - \I_0^*,
\ee
for $k=2$ we have local IV 
\be
 \IIstar - \I_0 - {\rm II} - {\rm IV} ,
\ee
and for $k=1$ we already have
\be
\IIstar - \I_0 - {\rm II} .
\ee
The new cycle $E_p$ supports $\I_0$ and we identified the intersection between $\I_0$ and $\I\I$ as a 5-brane.
\begin{figure}[t]
\begin{center}
\includegraphics{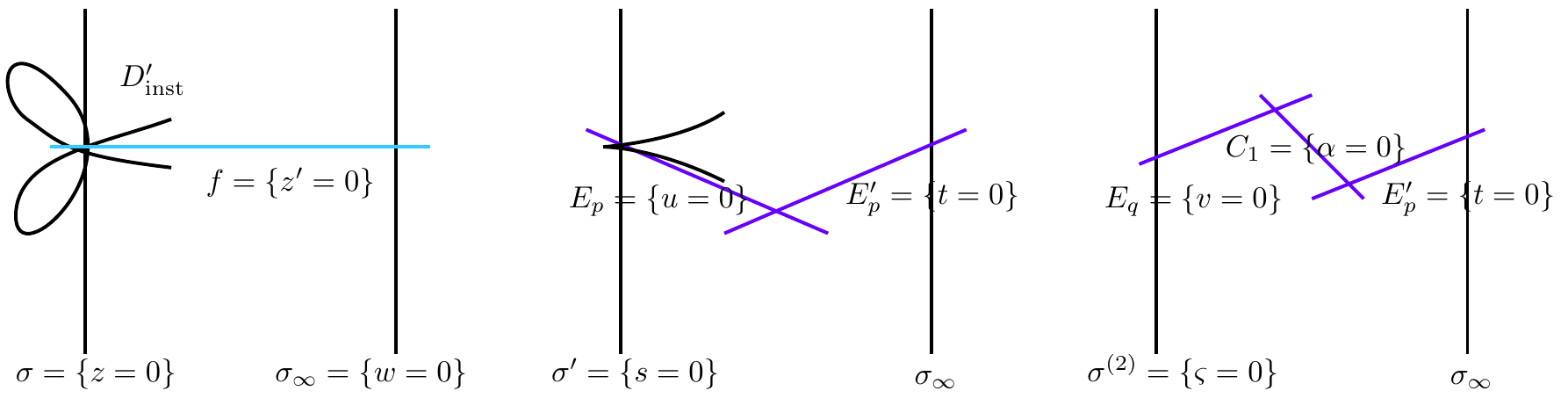}
\caption{\sl Resolutions of the $k=2$ case. We blow up the collision point between $\I\I^*$ and IV singularities to obtain $A_2$ configuration. Compare with Fig. \ref{f:A1}. \label{fig:IVresolution}}
\end{center}
\end{figure}

To sum up, we have two classes of SCFT coming from small instantons. One is from coalescent small instantons that we have seen here, and the other is from disjoint instantons by making 5-branes coincident that we have seen in the last section. The regular resolution chain is characterized by the relation to heterotic string, forming the M5 moduli space from the heterotic small instantons. The special resolution should be done for complete resolution, which is controlled by native IIB tensor multiplets, which ruins Calabi--Yau condition. There is no ambiguity in counting the number of 5-branes.

\section{Anomaly Cancellation} \label{s:anomaly}

Since 5-branes are induced from intersections between two 7-brane stacks, their types and numbers are completely determined by the arrangement of the 7-branes, which is encoded in the discriminant (\ref{disclocus}). In the global description, once the Calabi--Yau condition (\ref{CYcond}) is satisfied, the discriminant condition (\ref{disclocus}) guarantees
globally consistent and anomaly free vacuum. { We recapitulate its verification in Section \ref{s:anomaly}.1 by analyzing six-dimensional anomaly structures of  the 5-brane worldvolume. In Section \ref{s:anomaly}.2, we analyze the two-dimensional worldsheet anomalies of E(elementary)-strings, which provide us the strong evidences on the existence of \E-string with the $USp(q)$ gauge theory and the merging process of of M, G, H-strings.}

\subsection{Generalized Green--Schwarz mechanism}

For generic base $B$ of elliptic Calabi--Yau threefold, the number of tensor multiplets $n_T$ is given by the number of $(1,1)$-{cocycles} in the base surface $B$ \cite{Vafa:1996xn}
\be \label{NTcondition}
 n_{T} = h^{1,1}(B) - 1.
\ee
The subtraction by one unit is because one combination giving K\"ahler class belongs to the hypermultiplet, measuring the volume of $B$ \cite{Morrison:1996na}. 

First, we consider the six-dimensional theory. It is specified by tensor multiplets in which we have anti-self-dual two-forms $\omega^i$ in $H^{1,1}(B)$. Their intersections
\be \label{6Dlatticeintersec}
 \eta^{ij} = - \omega^i \cdot \omega^j, \quad a^i = - K_B \cdot \omega^i, \quad b_a^i = - \omega^i \cdot \sigma^a, 
\ee
naturally define an integral lattice.

With 7-branes, the ten-dimensional Bianchi identity for the self-dual five-form field strength $F_5$ of type IIB theory is modified as  \cite{Sadov:1996zm,Ohmori:2014kda}
\be \label{10DGS}
 d F_5 =  Z,
\ee
where  $Z$ reflects the contributions from 7-branes
\be \label{Z} \begin{split}
 Z =& \frac14 K_B \wedge p_1(T_6) + \sigma \wedge\frac12 \Tr F_{E_8, L}^2 + \sigma_\infty \wedge \frac12 \Tr F_{E_8, R}^2.
\end{split}
\ee
Here, $T_6$ is the six-dimensional tangent bundle, $\Tr$ denotes the trace over $\bf 248$ representation of $E_8$ divided by its dual Coxeter number $h_{E_8}^\vee= 30$, and the field strengths include both background and fluctuations. 
The $F_5$ in Eq.(\ref{10DGS}) is expanded by the above two-cocycles
$$ F_5 = H_i \wedge \omega^i, $$
yielding six-dimensional three form field strengths $H_i = dB_i${, under which strings are charged. We have scalar fields parameterizing the volumes of two-cycles. These form the bosonic components of the six-dimensional ${\cal N}=(1,0)$ tensor multiplets.}

The relation (\ref{10DGS}) now becomes six-dimensional Bianchi identities 
\be \label{6DGS}
 d H_i = I_i
\ee
with the 4-form polynomial $I_i$ accounting for the string from $\omega^i$ \cite{Ohmori:2014kda} 
\be 
 \eta^{ij} I_{j} =    - Z \cdot \omega^i  = \left(\frac14 a^i p_1(T_6) + \frac12 b^{i}_a\Tr F_{a}^2 \right),
\ee
where the intersection numbers are given in Eq.(\ref{6Dlatticeintersec}) for divisors $\sigma^a$ supporting singularities.
However, this is done up to a contribution from the six-dimensional ${\cal N} = (1,0)$ $R$-symmetry in Eq.(\ref{Rsymmetry6D}). This could not be obtained from the above IIB consideration, because the $R$-symmetry is local Lorentz symmetry in the heterotic or M-theory dual side. However, we can determine it from the six-dimensional anomaly structure  \cite{Ohmori:2014kda},
\be \label{I4inherited}
 \eta^{ij} I_j =  \left( \frac14  a^i p_1(T_6) +  \frac12 b^{i}_a\Tr F_{a}^2 \right) + h^{\vee}_{G_i} c_2(D).
\ee
Here, $h^{\vee}_{G_i}$ is the dual Coxeter number of the gauge group localized on the 7-brane wrapped on $\omega_i$ and $c_2(D)$ is the second Chern class for the $R$-symmetry $SU(2)_D$ bundle \cite{Ohmori:2014kda}. If we have no gauge group, we take $h^{\vee}_{\emptyset}=1$.
With these, we can check that six-dimensional Green--Schwarz conditions are satisfied \cite{Sadov:1996zm}.

The six-dimensional consistency condition for the lattice was studied in \cite{Seiberg:2011dr}. First, locality of surface operators involving the above $B_i$ requires that the lattice spanned by $\{ \omega_i \}$ be integral. Also, upon compactification down to two dimensions, the modular invariance of the resulting partition function constrains that the lattice should be unimodular
\be \label{unimodular}
 \det \eta = 1.
\ee

Throughout this work, we are taking the base $B=\F_n^{(k)}$ as a blown-up Hirzebruch surface $\F_n$ at $k$ small instanton points in $\sigma \cap D'_{\text{inst}}$ or $\sigma' \cap D'_{\text{inst}}$. $H^{1,1}(\F_n)$ is spanned by $\{f,\sigma\}$, so it includes $f$ associated with a tensor multiplet containing the heterotic dilaton. Each blow-up gives rise to an exceptional divisor $E_{p_i} \simeq \P^1 \in B$ and hence give one tensor multiplet. We also have seen that not every resolution of the base $B$, especially a blow-up at a point that is not these instanton points, can keep the vanishing first Chern class condition (\ref{CYcond}).

In our framework where all the strings are formed by E-strings, we may understand why the top-down approach naturally gives rise to the self-dual integral lattice.
In the limiting case where we resolve all the 24 small instanton points, we have as many E-string cycles $E_{p_i}$ and $E_{p_j} \sim f- E_{p_j}'$ forming a lattice $\Z^{24}$. Then it automatically satisfies the condition (\ref{unimodular}), 
$$  \eta^{ij} = - E_{p_i} \cdot E_{p_j} =  \delta^{ij}. $$
In the starting setup, specified by Eq.(\ref{smallinst}), we had in total 24 instantons counted by Eqs. (\ref{E8inst}) and (\ref{E8primeinst}). This can be regarded as a special case when $(12-n)$ cycles $E_{p_i}$ and $(12+n)$ cycles $E_{p_j}'$ shrink to have zero volume. 
This means that, if we generate the lattice $\{ \omega^i \}$ as linear transformation of $\{ E_{p_i} \}$  into $ADE$-lattice, we can guarantee the integral and self-dual condition. For example, if we take $A_2$ system by taking combinations $C_1 = E_{p_1} - E_{p_2},C_2=E_{p_2} - E_{p_3}$, we have a new basis $\{C_1, C_2, E_{p_3},\dots E_{p_{24}}\}$
\be \eta^{ij}=
\begin{pmatrix}
 2 & -1 & 0  & \\ -1 & 2 & -1 & \\ 0 & -1 & 1 & \\ & & & \ddots  
\end{pmatrix},
\ee
{with} the same determinant. 

The new intersection matrix becomes the charge of the string sourcing the two-form field in the local six-dimensional theory. Again, such string can be understood as a combination of E-strings. Also, the relative charge relation, encoded in $b_a^i$, give rise to gauge couplings \cite{Sadov:1996zm,Morrison:1996na,Bershadsky:1996nh}. 

Note that, the generator (\ref{obliquestring}) of $E_8$ is compatible with the condition, although it contains half-integral components.

\subsection{Anomalies on strings}

Strings carry their own anomalies on the worldsheet as well as, being defects, anomalies inflowed from higher-dimensional bulk. From the the anomaly structure we may verify that all the strings are composed of E-strings. 

The anomaly polynomial for a two-dimensional string from a wrapped D3-brane on two-cycle $\omega_i$ is given as \cite{Kim:2016foj,Shimizu:2016lbw}
\be \label{2Danomaly}
 I_4 = \eta^{ij} q_i \left(  \frac12 q_j \chi_4 (N) +  I_j \right),
\ee
where $I_j$ are the anomaly four-forms in Eq.(\ref{I4inherited}) appearing in the six-dimensional Bianchi identity (\ref{6DGS}), $\eta^{ij}$ is defined in Eq.(\ref{6Dlatticeintersec}), and $q_i$ are the charges of the strings. 
Also $\chi_4(N)=c_2(R) - c_2(L)$ is the Euler class of the normal bundle $N$ to the string in the M5-brane{, expressed in terms of the second Chern classes of the $SU(2)_R$ and $SU(2)_L$ bundles.} Moreover, $p_1(T_6)=p_1(T)+p_1(N)$ is the first Pontryagin class of the tangent $T$ and the normal $N$ bundles with respect to strings.

\subsubsection*{E-string}

First, we consider an emission of one point-like instanton at a base point $p$. This gives rise to an exceptional cycle $E_p$ and proper transform $f_p \sim E_p + E_p'$. 

Wrapping a D3-brane on this $E_p$ cycle $q$ times gives rise to an E-string, having charge $q$ under the dual two-form. The intersection number of $E_p$ is converted to $\eta^{pp} = - E_{p} \cdot E_{p} = 1$. Since we have no 7-brane wrapped on $E_p$ supporting gauge theory, we take the corresponding dual Coxeter number to be 1. We have
\be 
\begin{split}
 I_4^{qE} &=  \frac{q^2-q}{2} c_2 (L) -\frac{q^2+q}{2} c_2 (R)+ \frac{q}{2} \Tr F^2_{E_8,L} + \frac{q}{4} p_1 (T) - q c_2(G). \\
 \end{split}
\ee

This admits interpretation in terms of the worldsheet fields. We have $SU(2)_R \times SU(2)_G$ doublet $O(q)$ symmetric hypermultiplet of negative chirality and $SU(2)_L \times SU(2)_G$ doublet adjoint vector multiplet of positive chirality, explaining the multiplicities of the corresponding {characteristic classes}. Recall that here $SU(2)_G$ is the $R$-symmetry of six-dimensional ${\cal N} = (1,0)$ supersymmetry in the absence of the strings. The coefficient of $c_2(G)$ has contribution from the both of $ \frac{q^2-q}{2}-\frac{q^2 + q}{2} = -q$, as alluded below Eq. (\ref{Rsymmetry6D}). It is well-known that, for a single M2-brane $q=1$, there is no $SU(2)_L$ charged fermion, and this is verified by the coefficient of the $c_2(L)$. 

The $E_8$ charged Fermi supermultiplets explain the third term with $b^p_{\sigma'} = E_p \cdot \sigma' =1.$ The fourth is for gravitational anomaly from the tangent bundle $T$.

\subsubsection*{E$'$- and heterotic string}

We consider next the E$'$-string from a D3-brane wrapping $E_p'$ by $q'$ times. From the above discussion, we may exchange this string with the E-string from $E_p$ using the relation (\ref{Eprimeexchange}).  We may treat $E_p'$ as the exceptional divisor resulting from the blow-up on an instanton point on the right 7-brane $\sigma_\infty$. We have $\eta^{pp} = - E_p' \cdot E_p' =1$. The resulting anomaly polynomial reads
\be
 I_4^{q'E'}= - \frac{q^{\prime 2}+q'}{2} c_2(L)+\frac{q^{\prime 2}-q'}{2} c_2 (R) + \frac{q'}{2} \Tr F^2_{E_8,R} + \frac{q'}{4} p_1 (T) - q' c_2(D).
\ee
These are consistent with the fact that blow-down of $E$ increases the instanton number of the `left' $E_8$ counted by $\frac{q}{2} \Tr F^2_{E_8,L}$, while blowing down $E'$ increases that of the `right' $E_8$ counted by  $\frac{q'}{2} \Tr F^2_{E_8,R}$. 
It is known that $SU(2)_D$ should exist in the M-theory limit but invisible in the single M2-brane $q=1$ limit. This is true when we consider a single E-string but, on the other ``side'' of M5-brane connected to the the other M9, there indeed exists the group $SU(2)_D$. 

Then, we consider a combination of E-string and E$'$-string of the same charge $q' =q$
\be\begin{split}
 I_{4}^{qE} + I_{4}^{qE'} &= q \left (  \frac12 \Tr F_{E_8,L}^2 + \frac12 \Tr F_{E_8,R}^2 - c_2(T)-c_2(L)-c_2(R)-c_2(D)-c_2(G)  \right) \\
  & =  q \big ( c_2({\cal V}_{E_8 \times E_8}') - c_2({\cal T}_{10})   \big) =  I^{qf}_4 .
\end{split}
\ee
Here, we used the relation $p_1(T)=-2 c_2(T)$. The resulting anomaly is seen the same as that of the ten-dimensional heterotic string. On the first line, the first two terms combine into the second Chern class $ c_2({\cal V}_{E_8 \times E_8}')$ of the vector bundle of the $E_8\times E_8$. Using primed vector bundle, we highlight that a number of small instantons are now extracted. The last five terms form the second Chern class  $c_2({\cal T}_{10})$ of the entire ten-dimensional tangent bundle. This is understood as inflow from the 7-branes $\sigma$ and $\sigma_\infty$. Note that solely with this we cannot satisfy anomaly cancellation condition because some of small instantons are extracted into M5-branes.

\subsubsection*{$\bf {\tilde E}$ and M-string}

In the construction of $ADE$-type SCFTs, we met M-strings connecting 5-branes. They are regarded as either D3-branes wrapped on the cycles $C_i$ parameterizing the seperation of 5-branes or combinations of E and \E-strings. Here we verify this by considering anomaly structure.

A D3-brane wrapped on a cycle $-E_{p_{i+1}}$ by $q$ times gives rise to an \E-string with charge $-q$ under the dual tensor field $B_{i+1}$. We may also obtain it from E-string by the exchange in Eq.(\ref{minusEexchange}).  The resulting anomaly polynomial is
\begin{align} 
 I_4^{q\tilde E_{p_{i+1}}}&= \frac{q^2+q}{2} c_2 (L) -\frac{q^2-q}{2} c_2 (R)  - \frac{q}{2} \Tr F^2_{E_8,L} - \frac{q}{4} p_1 (T) + q c_2(D).
\end{align}
The sum reproduces the anomaly for the M-string
\be
 I_4^{qE_{p_i}} + I_4^{ q\tilde E_{p_{i+1}}} = q^2 ( c_2(L)-c_2(R) ) + q ( c_2(D)-c_2(G)).
\ee
The result shows that the anomaly arises from symmetric representation $q(q+1)/2$ of $O(q)$ and antisymmetric representation  $q(q-1)/2$ of $USp(q)$.
We have anomaly because the worldsheet theory is $(0,4)$ supersymmetric and hence still chiral.

As discussed in Sect. \ref{sec:inducedA1}, it is enhanced to a non-chiral adjoint fermion of $U(q)$ gauge theory in the local limit, where we have parity symmetric $(4,4)$ supersymmetry on the worldsheet, as shown in Eq.(\ref{Uqenhancement}). This is due to the unification with extra $(0,4)$ Fermi and hypermultiplets of opposite respective charities. 

Alternatively, we may consider a single cycle $C_1$ having self-intersection number $-2 = - \eta^{11}$ and du Val type $K_B \cdot C_1 = 0$ as in Eq.(\ref{DuVal}). If this cycle $C_1$ has no intersection with $\sigma$ and $\sigma'$, we obtain the same result from Eq.(\ref{2Danomaly})
$$
  I_4^{qM}  =  I_4^{qE_{p_i}} + I_4^{ q \tilde E_{p_{i+1}}}. 
$$
This is desired form because it becomes locally enhanced six-dimensional ${\cal N} = (2,0)$ symmetry. 

Note that the polynomial $I^{ \tilde E}$ is not equal to $-I^{E}$.  This is because \E-string is not the D3 wrapped on a cycle $-E$, which is  equivalent to anti-D3 wrapped on $E$. The remaining worldsheet of \E-string has opposite orientation to that of E-string. Therefore, even the corresponding the homological cycles sum up to zero, the corresponding anomaly polynomial does not vanish. However, we may check that a heterotic string is self-conjugate satisfying the relation $I^{-f} = I^f$.

\subsubsection*{G-string}

Finally, we compute the anomaly of G-string. This string is present in the global setup, as it is a D3-brane wrapped on the two-cycle $E_p - E_q' \sim E_p + E_q - f$. Thus the G-string is combination of E and \E$'$ string. 

The \E$'$-string of charge $q$ is defined as a D3-brane wrapped on $-E'$ by $q$ times and changing the orientation of the remaining direction. Thus it is the same BPS state as the E-string. They are converted by a successive application of the above two exchange operations in Eqs.(\ref{Eprimeexchange}) and (\ref{minusEexchange}). 

The \E$'$-string has the following anomaly polynomial
\begin{align} 
 I_4^{ q\tilde E'} &= - \frac{q^2-q}{2} c_2 (L) + \frac{q^2+q}{2} c_2 (R) - \frac{q}{2} \Tr F^2_{E_8,R} - \frac{q}{4} p_1 (T) + q c_2(G).
\end{align}
Hence we obtain the anomaly polynomial for the G-string by the sum
\be \label{Gstringpolynom}
I^{qG} = I_4^{qE} + I_4^{q\widetilde E'} = q  \left (\frac{1}{2} \Tr F^2_{E_8,L} -\frac{1}{2} \Tr F^2_{E_8,R}\right).
\ee
It does not care about the internal geometry but only the difference of the two vector bundles of $E_8$'s. The polynomial (\ref{Gstringpolynom}) is different from $I^{q E_p}+I^{qE_q} -I^f$.


We also checked other constituent strings that are used in $E_n$ type theory, considered in Section \ref{sec:morestrings}, and confirmed similar anomaly structure.

\section{Constructing Vacua}

The considerations we spelled out so far guide us to an important application: construction of six-dimensional vacua of non-perturbative heterotic string involving NS5-branes. First of all, the gauge groups are determined by configuration of 7-branes by choosing discriminant loci $D_a$ by the procedure described in Section \ref{s:sevenbranes}. Although in this paper we have focused on the limit where all the instantons shrink into points so that we have the full $E_8 \times E_8$ gauge symmetry, we may readily obtain smaller unbroken groups for realistic vacua by Higgsing. 

Recall that the small instantons of the heterotic string correspond to the intersection points between the divisors $D'_{\text{inst}}$ in Eq.(\ref{instantonic7}) and $\sigma$ (or $\sigma_\infty$). By adjusting $D'_{\text{inst}}$, we can modify these points. Also by blowing up some of these points, we induce 5-branes that are dual to NS5-branes in the heterotic side. Two-cycles from the blow-up are responsible for separation of 5-branes. Depending on combinations of cycles that D3-branes wrap, we may obtain various strings. Finally, E-strings are elementary in the sense that other strings like M-, G- and heterotic strings are obtained as their linear combinations. Using these building blocks, we may understand how the six-dimensional ${\cal N} = (2,0)$ and (1,0) SCFT sectors are embedded into F-theory and controlled by E-strings.

Each vacuum consists of two 7-branes supporting $E_8$'s at the opposite poles of $f \simeq \P^1$ in the base $B$ and a number of 5-branes in between. The novelty comes from the fact that, unlike the naive M-theory geometry where M9-branes at two ends of the interval $I$ and a number of M5-branes linearly aligned in between, the configuration of M5-branes have rich geometrical structure described by blow-ups in the base $B$. Viewing $f$ as a circle fibration over an interval $I$ and $T$-dualizing the circle fiber, we can go to the M-theory picture . 

One important consequence is the global consistency condition. We have just seen that how the anomalies of various strings can be decomposed into anomalies of elementary E-strings. The analysis in the last section shows that the following conditions are sufficient to have anomaly free vacuum.
\begin{enumerate}
\item The sum of the cycles $\omega_I$ wrapped by D3-branes is equal to that of heterotic string $f$
\begin{equation} \label{completeroute}
 \sum_I \omega_I  = f,
\end{equation}
where $\omega_I \in H^{1,1}(B)$ form an integral and self-dual lattice, spanned by $\{E_{p_i}, f\}$ with half-integral coefficients.
\item In Eq.(\ref{completeroute}), the index $I$ covers all the possible configuration of E-strings, whose number is $n_T$, the number of tensor multiplets coming from the blow-up of the small instanton points in the regular resolution chains. In other words, the path should pass through all the 5-branes, or the emitted small instantons of two 7-branes $\sigma$ and $\sigma_\infty$.
 \end{enumerate}
If a vacuum meets these conditions, the overall sum of two-dimensional anomaly-free condition reproduces the Bianchi identity for the three-form field strength of the heterotic string associated with $f$
\be \label{total10DBianchi}
 dH = q^2 \chi_4(N)+ q \big( n_{T} \chi_4(R)  - c_2 ({\cal V}_{E_8 \times E_8}') + c_2({\cal T}_{10})  \big)
\ee
in ten dimensions. The coefficient of $q$ is the well-known Bianchi identity for the three-form field strength $H$ containing $dB$, where $B$ is dual to $f$. This $B$ is mapped to the Kalb-Ramond two-form of heterotic string. The vector bundle $ {\cal V}_{E_8 \times E_8}'$ has missing contribution from the ejected NS5-branes. Namely, the Euler class $\chi_4(R)$ is naturally identified as Dirac delta function in the normal direction to the 5-branes, whose coordinates are collectively denoted as $x$ \cite{Freed:1998tg,Harvey:1998bx},
$$
 \delta^4 (x) d^4x = \chi_4 (R). 
$$
For the vanishing $q^2$ term in Eq.(\ref{total10DBianchi}), we need special condition $\chi_4(N)$ which is easily achieved in the flat space limit, that is ${\mathbb R}^4$ or four-torus $\mathbb{T}^4$. 

Interestingly, well-known constructions of six-dimensional ${\cal N} = (2,0)$ SCFT and E-string theories satisfying the SCFT block conditions  explicitly provide the local building block of vacua and solve the constraint (\ref{completeroute}). Hence, we can classify possible  globally allowed constructions. Viewing the configuration as collections of non-abelian cycles $C_J$'s considered in Sect. \ref{sec:nonabelian}, we may find locally gauge anomaly-free configuration. It is the theory of affine Lie algebra which classifies the possible linear combinations of integral lattice $E_{p_i}$ to solve the constraint
$$
 \sum a_J^\vee C_J = N,
$$
where $a_J^\vee$ are Dynkin indices including that of the extended root and $N$ is a null cycle. If this is part of regular resolution chain, the corresponding set of strings gives the 5-brane contribution proportional to $n_T$ in Eq.(\ref{total10DBianchi}). Classification of possible resolution has been made \cite{Heckman:2013pva,Heckman:2015bfa}. We may consider similar ones arising from the special resolution, in which we blow up in the bulk. However, we have to be careful about the globally valid embedding which should not ruin the Calabi--Yau condition (\ref{CYcond}). 

Gravitational anomaly can also be cancelled if the grand total anomaly is the same as that of ten-dimensional heterotic string. The contribution either comes from the M5-brane $\chi_4(R)$ or is inherited from the ten-dimensional tangent bundle ${\cal T}_{10}$. 
The total anomaly should not be changed under deformation, blow up and phase transition, as long as they are continuous. In general, configurations formed by combinations of E-strings are globally consistent. Thus, a consistent configuration is completely specified by the choice of Calabi--Yau manifold.

\subsection*{Acknowledgements} 
We are grateful to Jin-Beom Bae, Stefan Hohenegger, Amer Iqbal, Taro Kimura, Hiroyuki Shimizu, members of Fields, Gravity \& Strings $@$ IBS, and participants of ``Liouville, Integrability and Branes" (11) and (12) at Asia-Pacific Center for Theoretical Physics for stimulating discussions. The work of KSC was supported by the National Research Foundation of Korea (NRF) funded by the Ministry of Education under NRF-2015R1D1A1A01059940. 

\appendix

\section{Emission and absorption of 5-branes}

In this appendix, we provide algebraic descriptions on the emission, absorption and merging of 5-branes dual to M5-branes \cite{Aspinwall:1997ye,Gr,Mi}.

\subsubsection*{Blow up}

Consider one localized small instanton, as in Eq.(\ref{smallinstemitted}). It is located at $p=\{ z=z'=0\}$, which is the collision point of Kodaira $\IIstar$ at $z=0$ and $\I\I$ at $z'=0$. To blow up $\F^{(1)}_n \to \F_n$, we introduce a $\P^1$ whose homogeneous coordinates are $(s,t)$ such that,
\begin{equation} \label{blowingup}
 z = us, \quad z' = ut,
\end{equation}
and forbid $s=t=0$, that is, $st$ belongs to the Stanley--Reisner ideal $SR = \{ zw, z'w',st \}$. The original singular point $p$ has become an exceptional divisor $E_p=\{u=0\}$ of the resolved base $\F^{(1)}_n$. The new base coordinates are {$s,t,u$ that} have scaling \cite{Choi:2013hua}
\be
 (s,t,u) \sim (\nu s, \nu t, \nu^{-1} u), 
\ee
showing that $E_p$ is a $(-1)$-curve.
After the proper transform and rescaling, we have the Weierstrass equation for the Calabi--Yau threefold $X_3^{(1)} = {\mathbb T} \to \F_n^{(1)}$:
\begin{equation} \label{oneM5}
 y^2 = x^3 + \left( f_8 (ut,w') s^4 w^4 \right) x + g_{11-n}(ut,w') t s^5 w^7 + g_{12}(ut,w') s^6 w^6 + g_{12+n}(ut,w') u s^7 w^5.
\end{equation}
We see that now the location of the $E_8$ that was before $\sigma=\{z=0\}$ is described now by $\sigma' = \{s=0\}$. The situation is depicted in Fig. \ref{f:basicE}. Then, $s=0$ means $z=0$ but the converse does not hold. Nothing happened to the 7-brane for that $E_8$  except blow-up and we just have a more proper coordinate $s=0$ from there. Still, the remaining small instantons are inside that $\sigma'$, whose location is described by the coordinate $u$. 

The resulting manifold $X_3^{(1)}$ has no enhanced singularity at the intersection $s=u=0$, as we can see from Eq.(\ref{oneM5}), that is, no further vanishing of $s^5$ term at $u=0$. We have only one extra tensor multiplet from the blow-up associated with the new 5-brane, whose scalar component controls the volume of $E_p$. Also, after the proper transform, each term satisfies the Calabi--Yau condition (\ref{CYcond}) under the new canonical class.  

Substituting Eq.(\ref{blowingup}), we have new $SR$ elements $usw, utw',st$ but there is no intersection between $E_p$ and $\sigma'$, the first becoming $sw$ and $uw$. Therefore, the $SR$ now becomes
\be
 SR = \{ sw, uw, tw', st, uw' \}.
\ee

\subsubsection*{5-branes and E-strings}

By the blow-up (\ref{blowingup}), the fiber $f$ of $\F_n$ is proper transformed to $E'_p =  \{ t = 0 \}$. This plays the same role to $\sigma_\infty$ as what $E_p$ does to $\sigma$, as can be explicitly checked in the describing equation (\ref{oneM5}). We know that one 5-brane is emitted because we are left with ${11-n}$ small instantons from Eq.(\ref{smallinstemitted}). Therefore, the intersection 
$$E_p \cap  E'_p$$
as in Eq.(\ref{M5location}) should correspond to the M5-brane in the M-theory dual. That is, in the F-theory side, the M5-brane is {\em induced} as an intersection between two local 7-branes at $E_p$ and $E_p'$.
This $E_p'$ is not discriminant locus if we keep higher order terms $O(s^6)$ which do not contain $t$. The scalar part of the corresponding tensor multiplet parameterizes the volume of $E_p$. The volume of $E_p'$ is complementary; roughly it is the volume of the fiber minus the volume of $E$.

Let us pause to see how two $E_8$'s can communicate. 
On $E'_p$, we have $s\ne 0$, so we can work on the patch $s=1$. The original scaling means $(z,w) = (us,w) = (u,w) \sim (\mu u, \mu w) = (\mu z, \mu w)$. Thus, $(u,w)$ is homogeneous coordinate of this $E'_p$. This shows that the two coordinates describing the base and the fiber of $\F_n^{(1)}$ are related; for instance, on the patch $u \ne 0$, an affine coordinate in $E'_p$ is $(1,w/u)$.  A similar holds true on $E_p$. We can travel from $s=0$ to $t=0$ through the bridge $E_p$ and from $t=0$ to $w=0$ through $E_p'$. However, the 5-brane is not moving through these bridges along the Higgs branch. As it is located at the intersection (\ref{M5location}), it can move between two 7-branes $\sigma$ and $\sigma_\infty$ along the tensor branch; first blowing up $E_p$ and then blowing down $E'_p$. In the M-theory picture, the location of the M5-brane is a point. In the $\F_n^{(1)}$ base, its coordinate $(u,w)$ projects a point in the space $f$ whose homogeneous coordinate contains $w$. 

\subsubsection*{Blow down}

Since $E_p' \cdot E_p' = -1$, we can blow down the $E_p'$ to obtain a smooth surface. After scaling, Eq.(\ref{oneM5}) becomes
\begin{equation}
 y^2 = x^3 + \left( f_8 s^4 w^4 t^4 \right) x + g_{11-n} s^5 w^7 t^7 + g_{12} s^6 w^6 t^6 + g_{12+n} u s^7 w^5 t^6.
\end{equation}
It intersects $\sigma_\infty = \{ w=0\}$ and $E_p = \{u=0\}$ which absorbs $E_p'$. So, setting
\begin{equation}
 wt = v, \quad ut= z', 
\end{equation}
we have
\begin{equation}
 y^2 = x^3 + \left( f_8 s^4 v^4 \right) x + g_{11-n} (z',w') s^5 v^7 + g_{12} (z',w') s^6 v^6 + g_{12+n} (z',w') z' s^7 v^5.
\end{equation}

We may now rename 
\begin{equation}
 g_{12+n} (z',w')z' \equiv g_{13+n}(z',w')
\end{equation}
which means that the small instanton located at $z'=0$, without changing its coordinate, is absorbed by the other $E_8$, located at $v=0$. We recover the scaling relation $(s,v) \sim (\mu s, \mu v)$ forming a new Hirzebruch surface $\F_{n+1}$. This is the well-known `elementary transformation' \cite{GH}.

Further extraction of small instantons is straightforward. During the entire process, the number of 5-branes plus the the number of small instantons is preserved. It is of course because they are induced from the intersections of $D'_{\rm inst}$ and because the number of intersections are preserved during the process. This is directly connected to the Calabi--Yau condition and the anomaly cancellation condition in the F-theory.

\subsubsection*{Bringing two instantons together}

Instead of blowing-down, we may blow-up $\F^{(2)}_n \to \F^{(1)}_n$ another small instanton point $q$ at, say, $ z=az'+bw'=0$. This becomes the zero factor of $g_{12-n}$, or,
\be
 g_{11-n} = g_{10-n} (az' + bw'), \quad a \ne 0.
\ee
Blow-up is done by introducing another $\P^1$ 
\be
 s=v \varsigma, \quad aut+bw' = a v \tau, 
\ee
having scaling  $(\varsigma,\tau,v) \sim (\rho \varsigma,\rho \tau, \rho^{-1}v)$.
After proper transform, we have 
\begin{equation} \label{twoM5}
 y^2 = x^3 + \left( f_8 \varsigma^4 w^4 \right) x + a g_{10-n} t \tau \varsigma^5 w^7 + g_{12} \varsigma^6 w^6 + g_{12+n} u v \varsigma^7 w^5.
\end{equation}
The resulting situation is illustrated in Fig. \ref{f:doubleE}.
Note the exchange symmetry $u \leftrightarrow v, t \leftrightarrow \tau$, which is also reflected in Fig. \ref{f:doubleE}.
Further iterations for other small instantons shall give essentially the same form, and blowing down $E'$ type divisors can be identically done as before.

Now, we identify $C_1 = E_p - E_q$ with defining equation 
\be \label{alpha}
 \frac{u}{v} = \frac{v\tau-bw'}{a vt} \equiv  \frac{\alpha}{a}.
\ee
Unless  $b=0$, this equation has no information about $v=0$. Tuning one hypermultiplet can make $b=0$, and Eq.(\ref{alpha}) becomes a linear equivalence relation
\be
 E_p \sim E_q,
\ee
with a new coordinate $\alpha$ replacing $u$.  Also, $ \tau = a \alpha t$ implies that $E_p' \sim E_q'$. 
We have a situation analogous to having two parallel branes in the flat space. 

These two branes become coincident when we blow down the cycle $C_1$, so that we have enhanced symmetry
\begin{equation} \label{twoM5}
 y^2 = x^3 + \left( f_8  \varsigma^4 w^4 \right) x +g_{10-n} \alpha t^2 \varsigma^5 w^7 + g_{12} \varsigma^6 w^6 + g_{12+n} v^2 \alpha  \varsigma^7 w^5.
\end{equation}
Still, $t=0$ does not intersect  $\varsigma$. We have two coincident 5-branes located at the intersection $u=t=0$, making up locally six-dimensional  ${\cal N} = (2,0)$ theory of $A_1$ type. Originally, the coordinates scaled as
\be \label{scaling1}
 (v,t, \alpha) \sim (\rho^{-1} v, \nu t, \nu^{-1} \rho\alpha  ).
\ee
To shrink the cycle, we have $\rho=\nu^{-1}$, thus
\be \label{scaling2}
 (v,t, \alpha) \sim (\nu v, \nu t, \nu^{-2} \alpha  ).
\ee
This shows the $\Z_2$ orbifold structure; the exceptional divisor $\{\alpha=0\}$ is ${\cal O}(-2)$ bundle over the $\P^1$ base whose homogeneous coordinates are $v$ and $t$. Similarly, if we instead take $-C=E_p'-E_q$, we would have blown-up $\Z_2$ orbifold where $\{\alpha=0\}$ is a ${\cal O}(-2)$ bundle over the $\P^1 \ni (u,\tau)$.

\subsubsection*{Two coalescent instantons $k=2$}

Alternatively, we consider two small instantons coalescent at one point $z=z'=0$, or $k=2$ case in Eq.(\ref{tunedm5}),
$$
y^2 = x^3 + \left( f_8 z^4 w^4 \right) x + g_{10-n} z^5 (z')^2  w^7 + g_{12} z^6 w^6 + g_{12+n} z^7 w^5.
$$
The two singularities $\I\I^*$ at $\sigma=\{z=0\}$ and IV at $z'=0$ collide. The situation is depicted in the leftmost figure in Fig. \ref{fig:IVresolution}. We blow up at this point
$$
 z=u s, \quad z'=u t ,
$$
thus $(s, t,u) \sim (\nu s, \nu t ,\nu^{-1} u),st  \in SR.$
After the proper transform and rescaling, the equation becomes
\be
 y^2= x^3 + \left( f_8 s^4 w^4 \right) x + g_{10-n} u t^2 s^5  w^7 + g_{12} s^6 w^6 + g_{12+n} u s^7 w^5.
\ee
Still, we have colliding singularities $\I\I^*$ and II at $s=u=0$, as described in the middle figure of Fig. \ref{fig:IVresolution}. We blow it up once again
\be \label{newalpha}
 s = v \varsigma, \quad u = v \alpha,
\ee
and  $( \varsigma ,\alpha,v) \sim (\rho \varsigma, \rho \alpha, \rho^{-1} v),\varsigma \alpha \in SR.$
Its proper transform gives now manifold with canonical Kodaira singularities,
\begin{equation} \label{k2again}
 y^2 = x^3 + \left( f_8  \varsigma^4 w^4 \right) x +g_{10-n} \alpha  t^2 \varsigma^5 w^7 + g_{12} \varsigma^6 w^6 + g_{12+n} v^2 \alpha  \varsigma^7 w^5.
\end{equation}
Since $v$ and $w$ are disjoint, we have no enhanced singularity at $v=w=0$. 
This equation is identical to the previous one (\ref{twoM5}), where we have two distinct small instantons blown up and align 5-branes in a linearly equivalent way. It is crucial that the definition of $\alpha$ in Eq.(\ref{newalpha}) agrees with Eq.(\ref{alpha}). In other words, the procedures of coalescing 5-branes and blow-ups do commute, as shown in Sec. \ref{s:regularres}. Indeed, the coordinates scale as in Eqs. (\ref{scaling1}) and (\ref{scaling2}).

We may interpret that locally at $\varsigma \simeq 0$, we have 
$$ y^2 = x^3 +  \varsigma^5,   $$
with the coefficient $g_{10-n} \alpha t^2 w^7$ being locally constant, so that the $\IIstar$ singularity is not enhanced at $v=0$. Also, locally at $\alpha=t=0$, the equation (\ref{k2again}) looks as
$$ y^2 \simeq x^3 + \alpha t^2, $$
and should be interpreted as collision of $\I\I$ at $\alpha=0$ and IV at $t=0$. 

However, this local identification is relative, depending on which local point we are interested in. We can consider the intersection $v=\alpha=0$. Around $\varsigma \simeq 0$, the local equation 
$$ y^2 = x^3 +  \alpha  $$ 
tells us that there is $\I_0$ at $v=0$ and it agrees with the previous result. However, expanding around $w=0$, we have
$$ y^2 = x^3 +  \alpha v^2, $$
indicating that the local singularity at $v=0$ is IV.

Observe that none of the above local resolutions can modify the $g_{12} z^6 w^6$ term. This means that the blow-up of a local singularity in the bulk of the base cannot give rise to rescaling of $x,y$ to have (\ref{modificationL}), fulfilling the Calabi--Yau condition (\ref{CYcond}).

\section{Local versus global singularity}

In the main text, we only considered small instantons controlled by $D'_{\text{inst}}$. In those, we can only have locally enhanced singularity around $\sigma=\{z=0\}$. 
On the other hand, we may have globally extended singularity along $z'=0$ as follows 
\be \label{E8atzprime}
 y^2 = x^3 + \left( z^{\prime \ell} f_{8-\ell}  z^4 w^4 \right) x + z^{\prime k} \big( g_{12-n-k}  z^5  w^7 + g_{12-k}  z^6 w^6 + g_{12+n-k} z^7 w^5 \big),
\ee
with the discriminant
$$ 
 \Delta = z^{10} w^{10} z^{\prime 2k}  \left(4 f_{8-\ell}^3 z^{\prime 3\ell-2k} z^2 w^2 + 27 g_{12-n-k}^2 w^{4} + \dots \right) ,
$$
where we assume $3 \ell-2k \ge 0$.
Because this contains the factor $z^{\prime 2k} $, we have a globally valid discriminant locus of the singularity of $\ord(f,g,\Delta)=(\ge k,k,2k)$ at $z'=0.$  In other words, we have another stack of 7-branes at $z'=0$, intersecting both other II$^*$ at $z=0$ and $w=0$, respectively.

Noting that the K3 geometry in the heterotic side is described by $f_8$ and $g_{12}$, we may show that this K3 has the same type of orbifold singularity $\C^2/\Gamma_G$ at $z'=0$ \cite{Aspinwall:1997ye}. In this case, we can have more types of singularities displayed in Table \ref{t:smallinsts}. 

\begin{table}[t]
\begin{center}
\begin{tabular}{ccccc} \hline \hline
fiber & $f$ & $g$ &  \# inst \\ \hline
 $\I_{m}$ & $(-3 h_{4-m}^2 w^{\prime 2m}+ f_{8-m} z^{\prime m}) z^4 w^4$ & $g_{12-n-k} z^{\prime k} z^5w^7+ 2 h_{4-m}^3 w^{\prime 3m} z^6 w^6 (k \ge m) $ & $k $ \\
 II & $f_{8-\ell} z^{\prime \ell} z^4 w^4 \ (\ell\ge 1)$ & $g_{11-n} z^{\prime} z^5 w^7 $ & 1\\  
  IV & $f_{8-\ell}  z^{\prime \ell} z^4 w^4 \ (\ell \ge 2) $ & $g_{10-n} z^{\prime 2} z^5 w^7 $ & 2 \\  
     $\I_0^*$ & $f_{8-\ell}  z^{\prime \ell} z^4 w^4\ (\ell \ge 2)$ & $ g_{9-n}z^{\prime 3} z^5 w^7  $ & 3 \\   
$\I^*_{m}$ & $z^{\prime 2} (-3 h_{3-m}^2 w^{\prime 2m}+ f_{6-m} z^{\prime m}) z^4 w^4$ & $  z^{\prime 3} (g_{12-n-k} z^{\prime k-3} z^5 w^7 +  2 h_{3-m}^3 w^{\prime 3m} z^6 w^6)$ & $k$ \\
 $ \rm{IV^*}$ & $f_{8-\ell}  z^{\prime \ell} z^4 w^4\ (\ell \ge 3)$ & $g_{8-n} z^{\prime 4} z^5 w^7  $ & 4 \\ 
$\rm III^*$ & $f_5 z^{\prime 3} z^4 w^4$ & $ z^{\prime 5} (g_{12-n-k} z^{\prime k-5} z^5 w^7+g_{7}  z^6 w^6)$ & $k$ \\
 $ \I\I^*$ & $f_{8-\ell}  z^{\prime \ell} z^4 w^4 \ (\ell \ge 4)$ & $g_{7-n} z^{\prime 5} z^5 w^7 $ & 5 \\  
\hline
\end{tabular} \caption{Coalescent $k$ small instantons of $E_8$, given by (local) $ADE$ singularities at $z'=0$ \cite{Aspinwall:1997ye}. Compare with Table \ref{t:singularities}. Since we have tuned all the terms to have $z'=0$ as a discriminant locus, we have a globally valid 7-brane at $z'=0$.}
\label{t:smallinsts}
\end{center}
\end{table}


\end{document}